\documentclass[twocolumn]{article}
\usepackage[T1]{fontenc}
\usepackage[latin1]{inputenc}
\usepackage{graphics}
\usepackage{amsmath}
\usepackage[dvips]{graphicx}
\usepackage{amsfonts}
\usepackage{amssymb}

\begin{document}

\title{Ray travel times in range-dependent acoustic waveguides}
\author{\ Anatoly L. Virovlyansky\\\textit{{\small Institute of Applied Physics, Russian Academy of Science,}}\\\textit{{\small 46 Ul'yanov Street, 603950 Nizhny Novgorod, Russia}}}
\date{\ }
\maketitle
\begin{abstract}
\bigskip An approximate analytical approach to describe the stochastic motion
of sound rays in deep ocean is developed. This is done for a realistic
propagation model with an internal wave induced perturbation imposed on the
smooth background sound speed field. The chaotic ray dynamics is analyzed
using the Hamiltonian formalism taken in terms of the action-angle canonical
variables. It is shown that even at ranges of a few thousands km, the
magnitude of range variations of the action variable is still small enough to
be used as a small parameter in the problem. A simple expression for the
difference in travel times of perturbed and unperturbed rays and approximate
analytical solutions to stochastic ray (Hamilton) equations are derived. These
relations are applied to study range variations of the timefront (representing
ray arrivals in the time-depth plane). Estimations characterizing the widening
and bias of timefront segments in the presence of perturbations are obtained.
Qualitative and quantitative explanations are given to surprising stability of
early portions of timefronts observed in both numerical simulations and field experiments.
\end{abstract}

\section{\bigskip Introduction}

We consider the ray dynamics in a deep ocean acoustic waveguide with an
internal wave induced perturbation to the \ smooth (background) sound speed
field. It is assumed that statistics of random internal waves are determined
by the empirical Garrett-Munk spectrum \cite{FDMWZ79,CB98,BV98}. Numerical
simulations demonstrate that although this perturbation is weak, it gives rise
to a rather strong ray chaos \cite{BV98,SFW97}. In the presence of internal
waves ray trajectories exhibit extreme sensitivity to initial conditions and
at ranges of a few thousand km parameters of a ray trajectory with the given
starting depth and launch angle are practically unpredictable and can be
described only statistically. On the other hand, even under conditions of ray
chaos the arrival pattern retains some of its features observed in the
unperturbed waveguide. In particular, both numerical simulations
\cite{BV98,SFW97} and field experiments \cite{AET1,AET2} show that early
portions of \ arrival patterns formed by steep rays manifest surprising
stability. Due to this property, the ray travel time is the main signal
parameter in the underwater acoustics experiments from which inversions is
performed to reconstruct ocean temperature field \cite{MW79,SM91}.

Our main objective is to develop an approximate analytical approach for
description of the ray motion, including analysis of ray travel times, at very
long ranges. We argue that this goal can be accomplished in the scope of a
perturbation theory based on the Hamiltonian formalism in terms of the
action-angle variables \cite{LLmech,AZ91}. Range variations of the refractive
index in our propagation model are not adiabatic and we cannot use the
adiabatic invariance of the action variable. Nevertheless, it turns out that,
even at long ranges, the variance of the action variable is small enough to be
considered as a small parameter in the problem.

In the present paper we derive an approximate analytic relation for the
difference in travel times of two rays one of which propagates in the
perturbed waveguide and another one in the unperturbed waveguide. \ We also
obtain approximate solutions to the stochastic Hamilton (ray) equations for
the action and angle variables. These solutions are then applied to analyze
statistics of ray travel times at ranges up to 3000 km.

Our attention is focused on the so-called timefront representing ray arrivals
in the time-depth plane. We estimate the width of different branches
(segments) of the timefront in the presence of perturbation and establish the
criterion of nonoverlapping of neighboring segments. It is also shown that the
perturbation causes not only a widening (dispersion) of timefront segments but
some regular bias of the segments as well. The estimation of this bias is also
presented. Predictions made with our approximate analytical approach are
verified by comparison to results of numerical simulations.

Note, that in the present paper we consider only internal-wave-induced
perturbations. An important issue of travel time biases due to mesoscale
inhomogeneities (see, e.g., Ref. \cite{S85}) has not been broached here.

The paper is organized as follows. The ray (Hamilton) equations in terms of
the position-momentum variables, $(p,z)$, as well as an environmental model on
which we rely in this paper are presented in Sec. \ref{induced}. The emphasis
in this section is on discussion of properties of timefronts obtained by
numerical solution of the ray equations. The action-angle canonical variables,
$(I,\theta)$, are introduced in Sec. \ref{variables}. Section \ref{travel}
presents the derivation of our main relation for the difference in travel
times, $\Delta t$, of perturbed and unperturbed rays. In Sec. \ref{stochastic}
we deduce approximate stochastic equations governing fluctuations of action
and angle variables due to random inhomogeneities. Solving this equation
yields statistical characteristics of $I$ and $\theta$. In Sec.
\ref{small_parameters} these characteristics are used to analyze relative
magnitudes of different constituents of $\Delta t$ and significantly simplify
the expression for $\Delta t$ by neglecting small terms. Our final expression
for travel time variations is applied to investigation of the timefront
structure. This is done in Sec. \ref{structure}. Section \ref{adiabatic} is
concerned with generalization of the analytical approach to a more realistic
environmental model. Our results are summarized in the final section. In the
Appendix we shortly discuss how the transformation from $(p,z)$ to
$(I,\theta)$ variables can be performed numerically using a standard ray code.

\section{ Timefronts in the presence of internal-wave-induced
perturbation\label{induced}}

\subsection{Ray dynamics in terms of the Hamiltonian formalism}

Consider wave propagation in a two-dimensional medium with the coordinates $r$
(range) and $z$ (depth). It is assumed that the $z$-axis is directed downward
and the plane $z=0$ is the sea surface. The ray trajectory $z(r)$ is
determined by the sound speed field $c(r,z)$ and can be found from Fermat's
principle \cite{AZ91,BornWolf,Abdullaev,JKPS94} according to which the first
variation of the functional
\[
S=c_{r}\,\int\frac{ds}{c(r,z)}%
\]%
\begin{equation}
=\int dr\,n(r,z(r))\sqrt{1+\left(  \frac{dz}{dr}\right)  ^{2}},
\label{FermatH}%
\end{equation}
vanishes at the ray trajectory. Here $n(r,z)=c_{r}/c(r,z)$ is the refractive
index, $c_{r}$ is a reference sound speed, and $ds=dr\,\left(  1+\left(
dz/dr\right)  ^{2}\right)  ^{1/2}$ is the arc length. The functional $S$
represents the so-called eikonal and it is related to the ray travel time,
$t$, by%

\begin{equation}
t=S/c_{r}. \label{t}%
\end{equation}

Formally considering Eq. (\ref{FermatH}) as an action function of some
mechanical system with the $r$-variable playing the role of time, one can
apply the standard relations of classical mechanics
\cite{AZ91,BornWolf,Abdullaev}. This yields explicit expressions for the
momentum,
\begin{equation}
p=n\frac{dz/dr}{\sqrt{1+(dz/dr)^{2}}}, \label{p_dzdr}%
\end{equation}
and the Hamiltonian,%

\begin{equation}
H=-\sqrt{n^{2}-p^{2}}. \label{H_n_p}%
\end{equation}
Equations $p=n\,\sin\chi$ and $H=-n\cos\chi$ relate the momentum and the
Hamiltonian to the ray grazing angle, $\chi$ \cite{SFW97}.

Expression (\ref{FermatH}) for the eikonal can now be rewritten as
\begin{equation}
S=\int(pdz-Hdr). \label{SH}%
\end{equation}
Ray trajectories are governed by the Hamilton equations \cite{BV98,SFW97}%

\begin{equation}
\frac{dz}{dr}=\frac{\partial H}{\partial p}=-\frac{p}{H}=\frac{p}{\sqrt
{n^{2}-p^{2}}},\label{dzdrH}%
\end{equation}%
\begin{equation}
\frac{dp}{dr}=-\frac{\partial H}{\partial z}=\frac{n\partial n/\partial
z}{\sqrt{n^{2}-p^{2}}}.\label{dpdrH}%
\end{equation}

Equation (\ref{FermatH})-(\ref{dpdrH}) present the Hamiltonian formalism in
terms of the momentum-position canonical variables. Later on (in Sec.
\ref{variables}) we shall introduce another pair of canonical variables,
namely, the action-angle variables.

\subsection{Environmental model}

In what follows we shall consider a model of the sound speed field in the
form
\begin{equation}
c(r,z)=c_{0}(z)+\delta c(r,z), \label{crz0}%
\end{equation}
where $c_{0}(z)$ is a smooth (background) sound speed profile, and $\delta
c(r,z)$ is a range-dependent perturbation.

The unperturbed profile $c_{0}(z)$ used in our numerical simulation is typical
for deep water acoustic waveguides. It is shown in the left panel of Fig. 1.
The sound-channel axis, i.e. minimum of the sound speed profile, is located at
a depth of $0.738$ km.

\begin{figure}[ptb]
\begin{center}
\includegraphics[width=7cm,keepaspectratio=true]{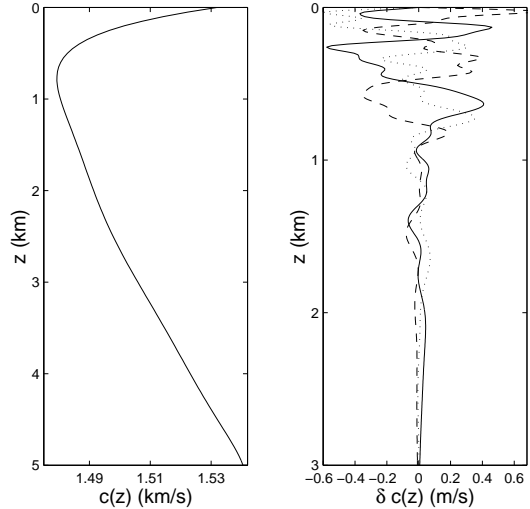}
\end{center}
\caption{Left panel: Background sound speed profile. Right panel: Sound speed
perturbation versus depth at three different ranges.}%
\label{ssp}%
\end{figure}

We consider an internal-wave-induced sound speed perturbation, $\delta c(r,z)
$, with a zero mean ($<\delta c(r,z)>=0$) and assume that the statistics of
the internal wave field is described by the empirical Garrett-Munk spectrum
\cite{FDMWZ79}. A numerical technique for generation of such perturbations
developed by J. Colosi and M. Brown \cite{CB98} has been applied. Equations
(1) and (19) from Ref. \cite{CB98} have been used to generate a particular
realization of $\delta c(r,z)$ which is used throughout this paper. It has
been assumed that the buoyancy frequency profile $\nu(z)$ is exponential,
$\nu(z)=\nu_{0}\exp(-z/B)$, and determined by two constants: a
surface-extrapolated buoyancy frequency $\nu_{0}=2\pi/10$ min$^{-1}$ and a
thermocline depth scale $B=1$ km. We consider the internal wave field formed
by $30$ normal modes and assume its horizontal isotropy. Components of wave
number vectors in the horizontal plane belong to the interval from $2\pi/100 $
km$^{-1}$ to $2\pi/4$ km$^{-1} $. An rms amplitude of the perturbation,
$(\delta c)_{\text{rms}}$, scales in depth like $\exp(-3z/2B)$ and its
surface-extrapolated value in our model is about $0.5$ m/s. Depth dependencies
of $\delta c$ at three different ranges are shown in the right panel of Fig. 1.

\begin{figure*}[ptb]
\begin{center}
\includegraphics[width=16cm,keepaspectratio=true]{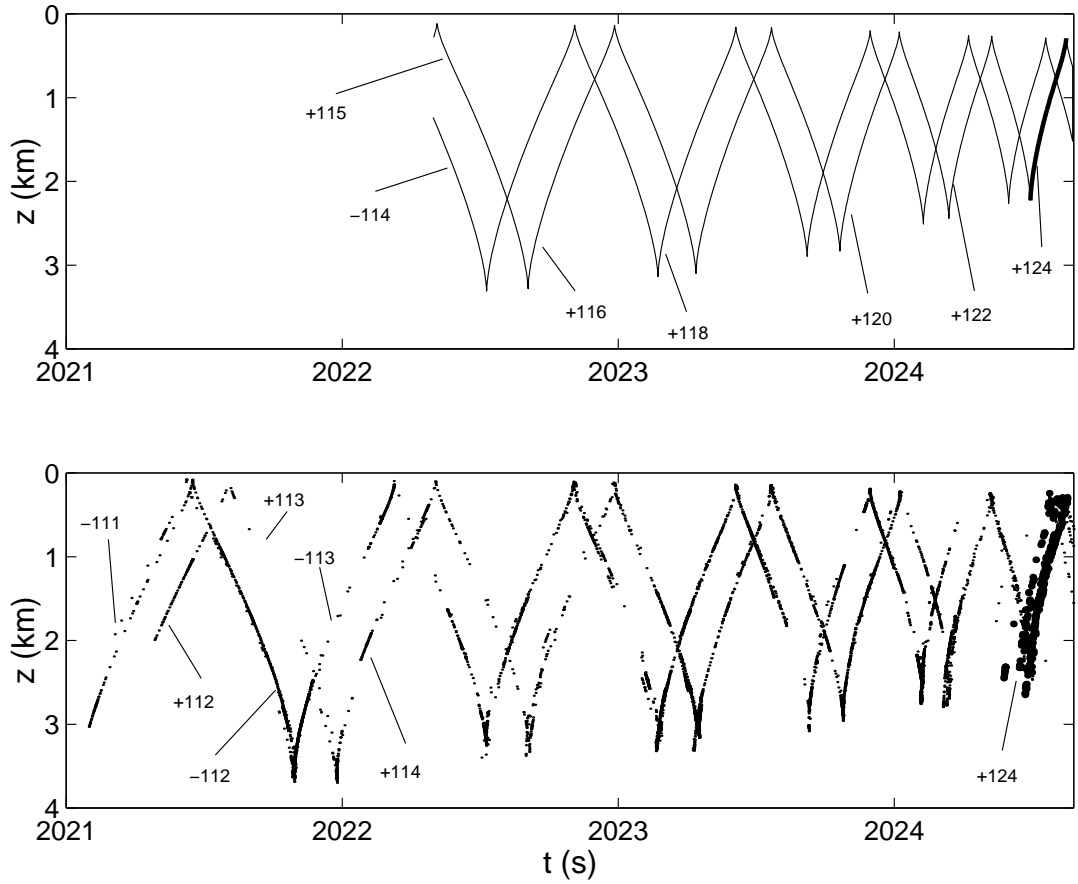}
\end{center}
\caption{Early portion of the timefront at the range 3000 km without (upper
panel) and with (lower panel) internal waves present. Identifiers of rays
forming some particular segments are indicated next to the corresponding
segments. In the upper panel, arrivals with identifier $+124$ are depicted by
a thick solid line. In the lower panel, arrivals with this identifier are
marked by thick points.}%
\label{tfront3000e}%
\end{figure*}

\begin{figure*}[ptb]
\begin{center}
\includegraphics[width=16cm,keepaspectratio=true]{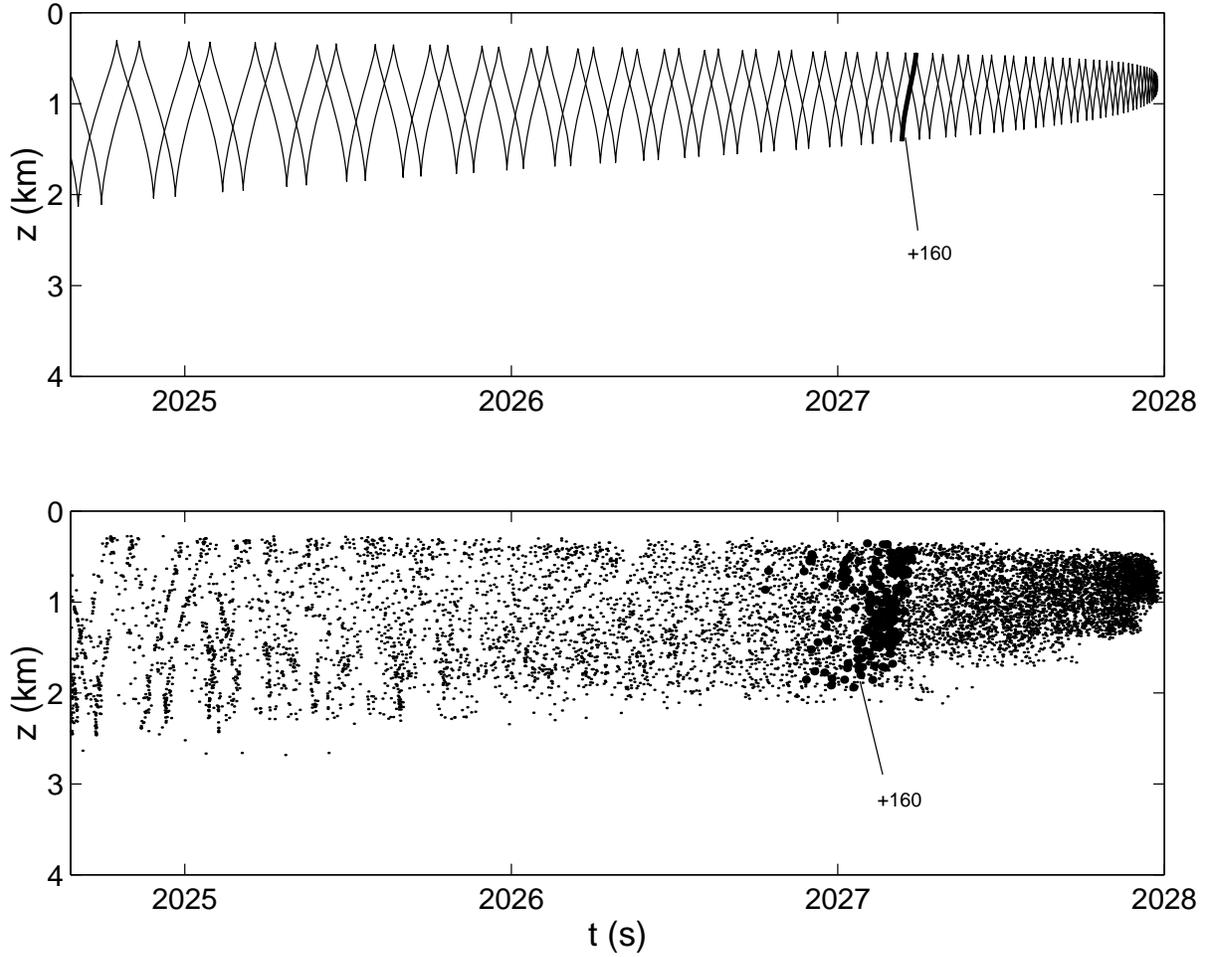}
\end{center}
\caption{Late portion of the timefront at the range 3000 km without (upper
panel) and with (lower panel) internal waves present. In the upper panel,
arrivals with identifier $+160$ are depicted by a thick solid line. In the
lower panel, arrivals with this identifier are marked by thick points.}%
\label{tfront3000l}%
\end{figure*}

\subsection{Numerical simulation of timefronts\label{timefront}}

Figures 2 and 3 show early and late portions of the timefront at 3000 km
ranges, respectively, for rays escaping a point source set at a depth of
$0.78$ km. The timefront in the unperturbed waveguide graphed in the upper
panels of Figs. 2 and 3 has been computed using a fan of $16000$ rays with
starting momenta equally spaced within an interval corresponding to launch
angles $\pm12^{\circ}$. The timefronts in the perturbed waveguide has been
produced by tracing $49000$ rays with starting momenta covering the
\textbf{same} interval.

The timefront in the range-independent waveguide has the well-known
accordion-like shape consisting of smooth segments (branches)
\cite{SFW97,BL91}. Each segment is formed by points corresponding to arrivals
of rays with the same identifier $\pm J$, where $J$ is the number of ray
turning points and symbols $+$ and $-$ correspond to rays starting upward and
downward, respectively. So, we can associate each segment with the identifier
of rays forming this segment. Identifiers for some particular segments in the
unperturbed waveguide are indicated in the upper panels of Figs. 2 and 3. It
is seen that the travel time grows with $J$. This is a typical situation for a
deep water waveguide \cite{BL91}: steep rays usually have greater cycle
lengths (smaller $J$) and arrive earlier than flat ones. Segments
corresponding to rays with launch angles of the same sign form a broken line.
Two such lines shifted along the $t$-axis form the unperturbed timefront (see
plots in the upper panels of Figs. 2 and 3).

A very interesting and important feature of the perturbed timefronts is a
remarkable stability of segments formed by early arriving steep rays. This
property of steep rays is well-known and it has been observed in both
numerical simulations and field experiments \cite{BV98,SFW97,AET1,AET2}.

\begin{figure}[ptb]
\begin{center}
\includegraphics[width=7cm,keepaspectratio=true]{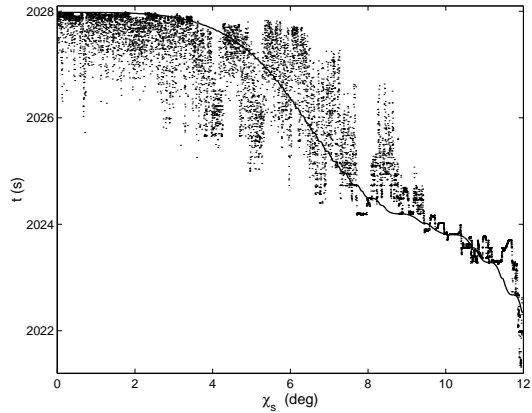}
\end{center}
\caption{Ray travel times at the range of 3000 km as a function of launch
angle, without (solid line) and with (points) internal waves present.}%
\label{tVSchi}%
\end{figure}

Figure 4 presents the ray travel time, $t$, as a function of $\chi_{s}$,
magnitude of the launch angle. Here and in the remainder of the paper we
present numerical results only for rays starting upward (for rays starting
downward the results are absolutely the same). In the range-independent case
the function $t(\chi_{s})$ is smooth and monotonous (solid curve). In
contrast, for the range-dependent case, we observe strong sensitivity of ray
travel time to starting angles: points depicting travel times of perturbed
rays are randomly scattered. Another manifestation of stochastic ray
instability in the presence of internal waves is seen in Fig. 5 where the ray
identifier at 3000 km range is shown as a function of the launch angle.
Perturbed rays with close initial conditions may have quite different number
of cycles. Although steep rays look less chaotic compared to flat ones (see
Fig. 4 and the upper panel in Fig. 5), a magnified view of the angular
interval $\chi_{s}>9^{\circ}$ shown in the lower panel of Fig. 5 demonstrates
that the dependence of $J$ on $\chi_{s}$ for steep rays in the perturbed
waveguide is also split into a set of irregular stripes.

\begin{figure}[ptb]
\begin{center}
\includegraphics[width=7cm,keepaspectratio=true]{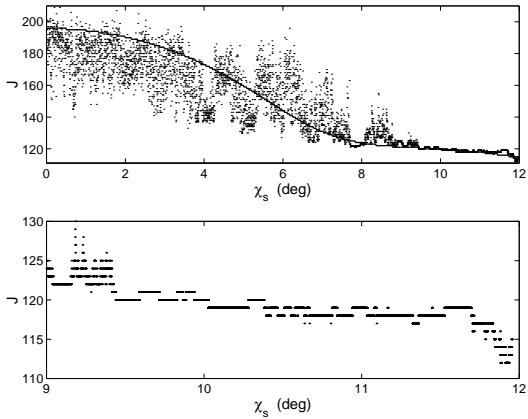}
\end{center}
\caption{Points in the top figure show the ray identifier at the range 3000 km
as a function of launch angle, in the presence of internal waves. The bottom
figure presents a magnified view of this dependence for steep rays taken
directly from the top figure. The solid line in the top figure graphs the
dependence $J(\chi_{s})$ in the unperturbed waveguide.}%
\label{idVSchi}%
\end{figure}

The time spread, $\delta t$, for rays with close launch angles in the
perturbed waveguide can be estimated as a width of the area occupied by
randomly scattered points in Fig. 4. For rays with $\chi_{s}<9^{\circ}$ this
yields $\delta t\approx1.5$ s. However, if we select a group of arrivals
corresponding to rays with the given identifier, we shall find that the time
spread of these arrivals will be significantly less than $\delta t$. It is
seen in Figs. 2 and 3 where two such groups of arrivals are shown by thick points.

So, in spite of strong sensitivity of ray parameters to initial conditions,
numerical simulations demonstrate an unexpectedly small time spread for rays
with the given identifier. Loosely, we can state that although the ray travel
time exhibits a chaotic and unpredictable dependence on the launch angle, its
dependence on the ray identifier is much more predictable. This effect is most
apparent for steep rays which form segments of the perturbed timefront almost
coinciding with the corresponding segments of the unperturbed timefront.
However, this fact does not mean that steep rays with the same launch angle in
the perturbed and unperturbed waveguides follows close ray paths.

To illustrate this statement and demonstrate that the situation is much more
reach and interesting come back to Fig. 2. Note that the perturbed timefront
begins earlier than the unperturbed one and six earliest segments in the lower
panel of Fig. 2 with identifiers $-111$, $\pm112$, $\pm113$, and $+114$ have
no counterparts in the upper panel. The point is that rays with these
identifiers in the unperturbed waveguide have launch angles exceeding the
maximum launch angle in our fan. In the presence of perturbation such rays
appear (see horizontal stripes with $J=112$, $113$, and $114$ in both panels
of Fig. 5 at $\chi_{s}$ close to $12^{\circ}$) because perturbation leads to
widening of the interval of ray grazing angles. As it has been indicated
already, in our waveguide steep rays arrive earlier than flat ones. So, it is
natural that in the presence of perturbation we observe rays arriving earlier
than those from the unperturbed fan. However, it is surprising and absolutely
unexpected that steep rays which appear due to scattering at random
inhomogeneities, form quite regular segments. Moreover, these segments
coincide with the unperturbed segments missed in the upper panel of Fig. 2.

One of our objectives is to explain how chaotic behavior of ray paths is
compatible with stability of early portions of the timefront formed by steep
rays and with unexpectedly small time spread of clusters of flat rays. Later
on, we shall address these issues using the Hamiltonian formalism in terms of
the action-angle canonical variables \cite{LLmech,AZ91}. The action-angle
variables are introduced in the next section.

\section{Action-angle variables\label{variables}}

\subsection{Range-independent waveguide}

First, we define the action-angle variables in a range-independent waveguide
with $c=c_{0}(z)$ and, correspondingly, $n=c_{r}/c_{0}(z)=n_{0}(z)$. In such a
waveguide the Hamiltonian remains constant along the ray trajectory:
\begin{equation}
H_{0}=-\sqrt{n_{0}^{2}(z)-p^{2}}.\label{SnellH}%
\end{equation}
This is the Snell's law \cite{BL91,JKPS94} (in geometrical optics it is often
presented in the form $n_{0}(z)\cos\chi=\text{const}$) analogous to the energy
conservation law in classical mechanics. Equation (\ref{SnellH}) establishes a
simple relation between the momentum $p$ and coordinate $z$
\begin{equation}
p=\pm\sqrt{n_{0}^{2}(z)-H_{0}^{2}}.\label{pEH}%
\end{equation}

The action variable $I$ is defined as the integral \cite{LLmech,AZ91}
\begin{equation}
I=\frac{1}{2\pi}\oint p\,dz=\frac{1}{\pi}\int_{z_{\min}}^{z_{\max}}%
dz\sqrt{n_{0}^{2}(z)-H_{0}^{2}} \label{IH}%
\end{equation}
running over the cycle of ray trajectory. Here $z_{\min}$ and $z_{\max}$ are
the depths of upper and lower ray turning points, respectively. Equation
(\ref{IH}) determines ``energy'', $H_{0}$, as a function of the action
variable, $I$. Note the relation
\begin{equation}
\frac{dH_{0}}{dI}=\frac{2\pi}{D}=\omega, \label{omegaH}%
\end{equation}
where
\begin{equation}
D=-2H_{0}\int_{z_{\min}}^{z_{\max}}\frac{dz}{\sqrt{n_{0}^{2}(z)-H_{0}^{2}}}
\label{DH}%
\end{equation}
is the cycle length of the ray path, and $\omega$ is the angular frequency of
spatial path oscillations. Equation (\ref{DH}) follows from (\ref{dzdrH}),
(\ref{pEH}), and (\ref{IH}).

The canonical transformation from the position-momentum, $(p,z)$, to the
action-angle, $(I,\theta)$, variables
\begin{equation}
I=I(p,z),\,\,\,\,\,\theta=\theta(p,z) \label{canonH}%
\end{equation}
and the inverse transformation
\begin{equation}
z=z(I,\theta),\quad p=p(I,\theta) \label{canoninvH}%
\end{equation}
are determined by the equation \cite{LLmech}
\begin{equation}
dS=pdz-H_{0}dr=dG-\theta dI-H_{0}dr, \label{dS}%
\end{equation}
where $G=G(I,z)$ is the generating function. An explicit expression for
$G(I,z)$ is well-known \cite{LLmech,AZ91}. We represent it in the form
\begin{equation}
G(I,z)=\left\{
\begin{array}
[c]{c}%
\int_{z_{\min}}^{z}dz\sqrt{n_{0}^{2}(z)-H_{0}^{2}(I)},\quad\;\quad p>0\\
2\pi I-\int_{z_{\min}}^{z}dz\sqrt{n_{0}^{2}(z)-H_{0}^{2}(I)},\quad p<0
\end{array}
\right.  , \label{GH}%
\end{equation}
where $z_{\min}$ and $z_{\max}$ are considered as functions of $I$. Then, equations%

\begin{equation}
p=\frac{\partial G}{\partial z}=\pm\sqrt{n_{0}^{2}(z)-H_{0}^{2}(I)},
\label{canon1H}%
\end{equation}
and
\[
\theta=\frac{\partial G}{\partial I}%
\]%
\begin{equation}
=\left\{
\begin{array}
[c]{c}%
\frac{2\pi H_{0}(I)}{D}\int_{z_{\min}}^{z}\frac{dz}{\sqrt{n_{0}^{2}%
(z)-H_{0}^{2}(I)}},\quad\;\quad p>0\\
2\pi-\frac{2\pi H_{0}(I)}{D}\int_{z_{\min}}^{z}\frac{dz}{\sqrt{n_{0}%
^{2}(z)-H_{0}^{2}(I)}},\quad p<0
\end{array}
\right.  \label{canon2H}%
\end{equation}
define the transformations (\ref{canonH}) and (\ref{canoninvH}).

Note, that the so defined angle variable $\theta$ varies from $0$ to $2\pi$ at
each cycle of the trajectory. We assume that the cycle begins at the minimum
of the trajectory. To make the angle variable continuous, its value should be
increased by $2\pi$ at the beginning of each new cycle. It should be
emphasized that both functions in Eq. (\ref{canoninvH}) are periodic in
$\theta$ with period $2\pi$.

The ray equations in the new variables take the trivial form
\begin{equation}
\frac{dI}{dr}=-\frac{\partial H_{0}}{\partial\theta}=0,\quad\frac{d\theta}%
{dr}=\frac{\partial H_{0}}{\partial I}=\omega(I) \label{dIdr0H}%
\end{equation}
with the solution
\begin{equation}
I=I_{s},\quad\theta=\theta_{s}+\omega(I_{s})\,r, \label{I0H}%
\end{equation}
where $I_{s}$ and $\theta_{s}$ are starting values of the action and angle
variables, respectively, at $r=0$.

Let us emphasize an almost trivial point which, nonetheless, is crucial for
our subsequent analysis. Although canonical transformations (\ref{canonH}) and
(\ref{canoninvH}) are determined by the function $n_{0}(z)$, formally, they
can be applied in a waveguide with a \textbf{different} refractive index
profile. Moreover, these transformations can be used in a range-dependent
waveguide, as well.

\subsection{Range-dependent waveguide\label{range-dep}}

Turn our attention to a more realistic model of the sound speed field given by
Eq. (\ref{crz0}). In this range-dependent environment with $n(r,z)=c_{r}%
/c(r,z)$ we define the action-angle variables using the same relations as in
the unperturbed waveguide with the refractive index $n_{0}(z)=c_{r}/c_{0}(z)$.

Rewrite the Hamiltonian $H=-\sqrt{n^{2}-p^{2}} $ in the form%

\begin{equation}
H=H_{0}+V, \label{HV}%
\end{equation}
with $H_{0}(p,z)$ defined by Eq. (\ref{SnellH}) and
\begin{equation}
V(p,z,r)=-\sqrt{n^{2}(z,r)-p^{2}}+\sqrt{n_{0}^{2}(z)-p^{2}.} \label{H0V}%
\end{equation}
If $|\delta c|/c_{r}<<1$, and $|p|<<1$ (in underwater acoustics both
conditions are typically met)
\begin{equation}
V=-\delta n, \label{V-delc}%
\end{equation}
where $\delta n=n-n_{0}\simeq\delta c/c_{r}$. Since our generating function
$G$ does not depend explicitly on $r$, the Hamiltonian in the new variables
$(I,\theta)$ is obtained by simple substitution of functions (\ref{canoninvH})
into the above equations. This yields
\begin{equation}
H(I,\theta,r)=H_{0}(I)+V(I,\theta,r). \label{HsH}%
\end{equation}

An explicit expression for the action function (eikonal) is obtained by
integrating Eq. (\ref{dS}) with $H_{0}$ replaced by $H$. We rewrite this
equation as
\begin{equation}
dS=d(G-\theta I)-Id\theta-Hdr \label{dS1}%
\end{equation}
and note that even though $G$ and $\theta$ are discontinuous at minima of the
ray trajectory, the term $G-\theta I$ vanishes at these points and, so, this
term varies continuously along the ray. The eikonal for a ray connecting
points $(0,z_{s})$ and $(r,z_{e})$ can be represented in the form
\[
S=\int\left(  Id\theta-Hdr\right)
\]%

\begin{equation}
-G(z_{s},I_{s})+G(z_{e},I_{e})+\theta_{s}I_{s}-\theta_{e}I_{e}, \label{S1H}%
\end{equation}
where $\theta_{s}$ and $\theta_{e}$ are angle variables at ranges $0$ and $r$,
respectively. They are defined by Eq. (\ref{canon2H}) with $z$, $I$, and $p$
being ray parameters at the beginning and at the end of the path. Note, that
the subscript '$s$' marks starting ray parameters while '$e$' marks parameters
at the end of the trajectory. This notation will be used throughout the paper.

In what follows we shall consider $\theta$ as a continuous variable defined in
accordance with the remark made after Eq. (\ref{canon2H}). Then
\begin{equation}
\theta(0)=\theta_{s},\quad\theta(r)=2\pi N+\theta_{e}-\theta_{s},
\label{theta}%
\end{equation}
where $N$ is the number of minima of the ray path, $\theta_{s}$ and
$\theta_{e}$ are the same quantities as in Eq. (\ref{S1H}) ($0\leq\theta
_{s},\theta_{e}\leq2\pi$), $\theta(0)$ and $\theta(r)$ are values of the
continuous angle variable at the beginning and at the end of the ray path, respectively.

The Hamilton equations now take the form%

\begin{equation}
\frac{dI}{dr}=-V_{\theta}, \label{dIdrH}%
\end{equation}
and
\begin{equation}
\frac{d\theta}{dr}=\omega+V_{I}, \label{dtedrH}%
\end{equation}
where
\[
V_{\theta}\equiv\frac{\partial V}{\partial\theta},\,\;V_{I}\equiv
\frac{\partial V}{\partial I}.
\]

Two comments should be made to this definition of the action-angle variables.

(i) The action variable introduced in this way (we have followed Refs.
\cite{AZ91,Abdullaev}) does not conserve along the ray path even in a
waveguide with very smooth range-dependence, i.e. our actions are not
adiabatic invariants. Another definition of these variables \cite{LLmech}
where the action does have a property of adiabatic invariance is shortly
described in Sec. \ref{adiabatic}.

(ii) Splitting of the Hamiltonian into a sum of the unperturbed constituent,
$H_{0}$, and the perturbation, $V$, have been made in anticipation of our
later use of a perturbation expansion based on smallness of $\delta c$ and,
hence, $V$. However, for now we have not assumed the perturbation to be small
and all equations derived so far are exact.

\section{Explicit expression for difference in ray travel times\label{travel}}

\bigskip In this section we compare two rays one of which propagates in an
unperturbed range-independent waveguide with $\delta c=0$, while another one
propagates in a range-dependent waveguide with nonzero $\delta c$. Both rays
start at $r=0$ and our task is to derive an analytical relation for the
difference in their travel times at a given range $r>0$. Starting parameters
of the rays are, generally, different but we assume that their action
variables remain more or less close at any intermediate range. This assumption
will be quantified later on.

To distinguish between similar parameters of the two rays under consideration,
parameters of a ray in the unperturbed waveguide will be marked with the
overbar. For example, starting and final parameters of its trajectory will be
denoted by $(\bar{p}_{s},\bar{z}_{s})$ and $(\bar{p}_{e},\bar{z}_{e})$,
respectively, while for another ray we shall write $(p_{s},z_{s})$ and
$(p_{e},z_{e})$.

The symbol $\Delta$ will be used to denote the difference between any
characteristic of one ray and its counterpart for another ray, e.g. $\Delta
I=I-\bar{I}$, $\Delta S=S-\bar{S}$, $\Delta z_{s}=z_{s}-\bar{z}_{s}$, and so on.

\subsection{Difference in eikonals}

Presenting Eq. (\ref{dtedrH}) in the form
\begin{equation}
d\theta=(\omega+V_{I})\,dr \label{d_theta}%
\end{equation}
and substituting it into Eq. (\ref{S1H}) we transform the expression for the
eikonal to
\begin{equation}
S=S_{G}+S_{0}+S_{V}, \label{S_split}%
\end{equation}
where
\[
S_{G}=-G(z_{s},I_{s})
\]%
\begin{equation}
+I_{s}\theta_{s}+G(z_{e},I_{e})-I_{e}\theta_{e}, \label{S_G}%
\end{equation}%
\begin{equation}
S_{0}=\int\left(  I\omega-H_{0}\right)  \,dr, \label{S_0}%
\end{equation}%
\begin{equation}
S_{V}=\int\left(  IV_{I}-V\right)  \,dr. \label{S_V}%
\end{equation}

Equations (\ref{theta}) and (\ref{d_theta}) yield
\begin{equation}
2\pi N+\theta_{e}-\theta_{s}=\int\left(  \omega+V_{I}\right)  \,dr.
\label{Del_theta}%
\end{equation}
For a ray in the unperturbed waveguide ($V=0)$ the action $\bar{I}$ does not
depend on range and Eqs. (\ref{S_split})--(\ref{Del_theta}) translate to
\begin{equation}
\bar{S}=\bar{S}_{G}+\bar{S}_{0}, \label{Sbar_split}%
\end{equation}%
\begin{equation}
\bar{S}_{G}=-G(\bar{z}_{s},\bar{I})+G(\bar{z}_{e},\bar{I})+(\bar{\theta}%
_{s}-\bar{\theta}_{e})\bar{I}, \label{S_GU}%
\end{equation}%
\begin{equation}
\bar{S}_{0}=\int\left(  \bar{I}\bar{\omega}-\bar{H}_{0}\right)  \,dr,
\label{S_0U}%
\end{equation}%
\begin{equation}
2\pi\bar{N}+\bar{\theta}_{e}-\bar{\theta}_{s}=\int\bar{\omega}\,dr.
\label{Del_thetaU}%
\end{equation}

Turn our attention to the difference in eikonals
\[
\Delta S=S-\bar{S}%
\]%

\begin{equation}
=S_{G}-\bar{S}_{G}+S_{0}-\bar{S}_{0}+S_{V}. \label{Del_Sa}%
\end{equation}
Subtracting Eq. (\ref{S_GU}) from Eq. (\ref{S_G}), exploiting Eq.
(\ref{canon2H}), and retaining only the first order terms in $\Delta I$, we
get
\[
S_{G}-\bar{S}_{G}=-G(z_{s},\bar{I})+G(\bar{z}_{s},\bar{I})+G(z_{e},\bar
{I})-G(\bar{z}_{e},\bar{I})
\]%
\begin{equation}
+\bar{I}\left(  \Delta\theta_{s}-\Delta\theta_{e}\right)  . \label{Del-SG}%
\end{equation}
Represent the difference $S_{0}-\bar{S}_{0}$ as
\begin{equation}
S_{0}-\bar{S}_{0}=\int\left[  F(I)-F(\bar{I})\right]  \,dr, \label{Del-S0}%
\end{equation}
where
\begin{equation}
F(I)=I\omega(I)-H_{0}. \label{F}%
\end{equation}
Using Eq. (\ref{omegaH}), it follows that
\begin{equation}
F(I)-F(\bar{I})=\bar{I}\bar{\omega}^{\prime}+\sum_{\nu=2}^{\infty}\frac
{\omega^{(\nu-1)}+\bar{I}\omega^{(\nu)}}{\nu!}\Delta I^{\nu},
\label{Fexpansion}%
\end{equation}
where
\begin{equation}
\omega^{\prime}=\frac{d\omega(\bar{I})}{d\bar{I}},\;\omega^{(\nu)}%
=\frac{d^{\nu}\omega(\bar{I})}{d\bar{I}^{\nu}}. \label{omega_prime}%
\end{equation}

Subtracting Eq. (\ref{Del_thetaU}) from Eq. (\ref{Del_theta}) we get
\[
2\pi\Delta N+\Delta\theta_{e}-\Delta\theta_{s}=\int\left[  \omega
(I)-\omega(\bar{I})+V_{I}\right]  \,dr
\]%
\begin{equation}
=\int\left(  \sum_{\nu=1}^{\infty}\frac{\omega^{(\nu)}}{\nu!}\Delta I^{\nu
}\right)  dr+\int V_{I}\,dr. \label{Del-angle}%
\end{equation}
Multiplying this equation by $\bar{I}$ and combining it with Eqs.
(\ref{Del-SG}), (\ref{Del-S0}), and (\ref{Fexpansion}) we, finally, obtain
\[
\Delta S=2\pi\Delta N\,\bar{I}
\]%
\[
-G(z_{s},\bar{I})+G(\bar{z}_{s},\bar{I})+G(z_{e},\bar{I})-G(\bar{z}_{e}%
,\bar{I})
\]%
\[
+\int\left(  \sum_{\nu=2}^{\infty}\frac{\omega^{(\nu-1)}}{\nu!}\Delta I^{\nu
}\right)  dr
\]%
\begin{equation}
+\int\left(  \Delta I\,V_{I}\,-V\right)  \,dr. \label{Del-Sb}%
\end{equation}

\subsection{Constituents of travel time variation}

The difference in travel times of our two rays, $\Delta t=\Delta S/c_{r}$, can
be represented as a sum of 4 constituents:%

\begin{equation}
\Delta t=\Delta t_{G}+\Delta t_{N}+\Delta t_{I}+\Delta t_{V},
\label{constituents}%
\end{equation}
where
\begin{equation}
c_{r}\Delta t_{G}=-G(z_{s},\bar{I})+G(\bar{z}_{s},\bar{I})+G(z_{e},\bar
{I})-G(\bar{z}_{e},\bar{I}) \label{Del_t_G}%
\end{equation}%
\begin{equation}
c_{r}\Delta t_{N}=2\pi\,\Delta N\,\bar{I}, \label{Del_t_N}%
\end{equation}%
\[
c_{r}\Delta t_{I}=\,\int\left(  \frac{1}{2}\omega^{\prime}\Delta I^{2}%
+\frac{1}{3}\omega^{\prime\prime}\,\Delta I^{3}\right.
\]%
\begin{equation}
\left.  +\frac{1}{8}\omega^{\prime\prime\prime}\Delta I^{4}+\ldots\right)
\,\,dr, \label{Del_t_I}%
\end{equation}%
\begin{equation}
c_{r}\Delta t_{V}=\int\left(  \Delta IV_{I}-V\right)  \,dr. \label{Del_t_V}%
\end{equation}

In the next section we obtain approximate solutions to stochastic ray
equations in terms of action and angle variables. These solution provide
simple estimations of magnitudes of individual terms present in Eqs.
(\ref{Del_t_I}) and (\ref{Del_t_V}). Then the expression for $\Delta t$ can be
simplified by neglecting small terms. This is done in Sec.
\ref{small_parameters}.

\section{Stochastic ray dynamics in terms of action-angle
variables\label{stochastic}}

In this section proceeding from the Hamilton equations (\ref{dIdrH}) and
(\ref{dtedrH}) we develop a simple approximate statistical description of ray
trajectory fluctuations. This result is applied to estimate magnitudes of
terms present in Eqs. (\ref{Del_t_I}) and (\ref{Del_t_V}) and find small
parameters in the problem that may be used to simplify the expression for
$\Delta t$.

\subsection{Fokker-Planck equation for the distribution of
action\label{sec-Fokker}}

Consider the perturbation $V(I,\theta,r)$ as a random function with given
statistical characteristics. Then the Hamilton equations (\ref{dIdrH}) and
(\ref{dtedrH}) constitute a system of stochastic (Langevin) equations. We
assume that the medium inhomogeneities are so weak that the horizontal scale
of their fluctuations, $l_{m}$, is much less than the scale of range
variations of the action, $l_{I}$,
\begin{equation}
l_{m}\ll l_{I}. \label{lmlI}%
\end{equation}
This assumption allows one to derive the Fokker-Planck equation for the
probability density function (PDF) of the action $I$. Let us derive this
equation using the method described in Ref. {\cite{GMS91}}. Consider an
arbitrary smooth function $\phi(I)$ and derive an equation for $<\phi>$, where
the symbol $<...>$ means statistical averaging over random medium
inhomogeneities. Exploiting Eq. (\ref{dIdrH}) we get
\begin{equation}
\frac{d}{dr}<\phi>=-<\phi_{I}(I)\,V_{\theta}(I,\theta,r)>. \label{dfidr}%
\end{equation}
Here and in what follows the subscripts $I$ and $\theta$ at $V$ denote partial
derivatives of function $V(I,\theta,r)$ with respect to the first and second
arguments. Transform the right hand side of Eq. (\ref{dfidr}) exploiting the
condition (\ref{lmlI}). The latter ensures that there exists a range $r_{0}<r$
such that
\begin{equation}
l_{m}<r-r_{0}<l_{I}. \label{lrl}%
\end{equation}
Denote the values of the angle and action variable at the range $r_{0}$ by
$\theta_{0}$ and $I_{0}$, respectively. These variables at the range $r$ may
be represented in the form
\begin{equation}
I=I^{(0)}+I^{(1)},\,\,\,\,\,\,\,\,\theta=\theta^{(0)}+\theta^{(1)},
\label{I0-T0}%
\end{equation}
where $I^{(0)}$ and $\theta^{(0)}$~are solution of Eqs. (\ref{dIdrH}) and
(\ref{dtedrH}) with initial conditions $I^{(0)}(r_{0})=I_{0}$ and
$\theta^{(0)}(r_{0})=\theta_{0}$. It is clear that
\[
I^{(0)}=I_{0},\,\,\,\,\,\,\,\theta^{(0)}=\theta_{0}+\omega^{\prime}%
(I_{0})\,(r-r_{0}).
\]
The symbols $I^{(1)}$ and $\theta^{(1)}$ denote first order corrections due to
the small perturbation $V$. By definition $I^{(1)}(r_{0})=0$ and $\theta
^{(1)}(r_{0})=0$. Then
\begin{equation}
I^{(1)}=-\int_{r_{0}}^{r}V_{\theta}^{(0)}(r^{\prime})dr^{\prime}. \label{I-1}%
\end{equation}
The superscript $(0)$ at $V_{\theta}$ as well as at other partial derivatives
of $V$ and at $V$ itself (see below) means that the arguments $I$ and $\theta$
of the corresponding function should be replaced with $I^{(0)}$ and
$\theta^{(0)}$, respectively. For example,
\[
V^{(0)}(r)\equiv V(I_{0},\theta_{0}+\omega^{\prime}(I_{0})\,(r-r_{0}),r).
\]
Using this notation we get
\[
\theta^{(1)}=\omega^{\prime}(I_{0})\int_{r_{0}}^{r}I^{(1)}(r^{\prime
})\,dr^{\prime}+\int_{r_{0}}^{r}dr^{\prime}\,V_{I}^{(0)}(r^{\prime}).
\]
After simple algebra the above expression may be presented in the form
\begin{equation}
\theta^{(1)}=I^{(1)}\frac{\partial}{\partial I_{0}}\theta^{(0)}+\frac
{\partial}{\partial I_{0}}\int_{r_{0}}^{r}dr^{\prime}\,V^{(0)}(r^{\prime}).
\label{T-1m}%
\end{equation}
Here we have exploited the relation
\begin{equation}
\frac{\partial V^{(0)}}{\partial I_{0}}=V_{I}^{(0)}+V_{\theta}^{(0)}%
\frac{\partial\theta^{(0)}}{\partial I_{0}}. \label{dV0dI0}%
\end{equation}

The statistical averaging denoted by the symbol $<>$ can be considered as an
averaging over ray parameters at the range $r_{0}$, i.e. over $I_{0}$,
$\theta_{0}$, and over medium inhomogeneities $V$ located within the range
interval $(r_{0},r)$. Let us present the term on the right of Eq.
(\ref{dfidr}) as
\[
<<\phi_{I}(I)\,V_{\theta}(I,\theta,r)>_{\theta_{0},V}>_{I_{0}}%
\]
where the inner brackets denote averaging over $\theta_{0}$ and the
inhomogeneities $V$, while the outer brackets denote averaging over $I_{0}$.
Let us first consider the conditional average denoted by the inner brackets.
Using the smallness of $I^{(1)}(r)$ and $\theta^{(1)}(r)$, which are terms
$O(V)$, we can simplify the conditional average
\[
<\phi_{I}(I)\,V_{\theta}(I,\theta,r)>_{\theta_{0},V}=<\phi_{I}\left(
I^{(0)}+I^{(1)}\right)
\]%
\[
\times\,V_{\theta}\left(  I^{(0)}+I^{(1)},\theta^{(0)}+\theta^{(1)},r\right)
>_{\theta_{0},V}%
\]%
\[
=\phi_{II}(I_{0})<I^{(1)}V_{\theta}^{(0)}>_{\theta_{0},V}%
\]%
\begin{equation}
+\phi_{I}(I_{0})<V_{\theta I}^{(0)}I^{(1)}+V_{\theta\theta}^{(0)}\theta
^{(1)}>_{\theta_{0},V}. \label{fiV}%
\end{equation}
The above expression provides an approximation to the right hand side of Eq.
(\ref{dfidr}) accurate up to terms $O(V^{2})$. Using Eq. (\ref{I-1}) we get
\begin{equation}
<I^{(1)}V_{\theta}^{(0)}>_{\theta_{0},V}=-\frac{1}{2}\frac{d}{dr}<\left(
I^{(1)}\right)  ^{2}>_{\theta_{0},V}. \label{I-1V}%
\end{equation}
Substituting this into the last average on the right of Eq. (\ref{fiV}) and
using Eq. (\ref{T-1m}) we obtain the following expression
\[
<V_{\theta I}^{(0)}I^{(1)}+V_{\theta\theta}^{(0)}\theta^{(1)}>_{\theta_{0}%
,V}=<I^{(1)}\frac{\partial}{\partial I_{0}}V_{\theta}^{(0)}%
\]%
\begin{equation}
+V_{\theta\theta}^{(0)}\frac{\partial}{\partial I_{0}}\int_{r_{0}}%
^{r}dr^{\prime}\,V^{(0)}(r^{\prime})>_{\theta_{0},V}. \label{av1}%
\end{equation}
We assume that the angle variable $\theta_{0}$ is uniformly distributed over
the interval $(0,2\pi)$. It means that the statistical averaging over
$\theta_{0}$ is defined by
\[
<...>_{\theta_{0}}=\frac{1}{2\pi}\int_{0}^{2\pi}...\,d\theta_{0}.
\]
This yields
\[
<\frac{\partial^{2}V^{(0)}(r)}{\partial\theta_{0}^{2}}\,V^{(0)}(r^{\prime
})>_{\theta_{0},V}%
\]%
\begin{equation}
=-<\frac{\partial V^{(0)}(r)}{\partial\theta_{0}}\,\frac{\partial
V^{(0)}(r^{\prime})}{\partial\theta_{0}}>_{\theta_{0},V} \label{uniform}%
\end{equation}
Using relations
\begin{equation}
V_{\theta}^{(0)}=\frac{\partial}{\partial\theta_{0}}V^{(0)}%
,\,\,\,\,\,V_{\theta\theta}^{(0)}=\frac{\partial^{2}}{\partial\theta_{0}^{2}%
}V^{(0)} \label{dTdT0}%
\end{equation}
(note that the partial derivative with respect to $I_{0}$ has not this
property: according Eq. (\ref{dV0dI0}) $V_{I}^{(0)}\neq\frac{\partial
}{\partial I_{0}}V^{(0)}$) we obtain
\[
<V_{\theta I}^{(0)}I^{(1)}+V_{\theta\theta}^{(0)}\theta^{(1)}>_{\theta_{0}%
,V}=\frac{\partial}{\partial I_{0}}<I^{(1)}V_{\theta}^{(0)}>_{\theta_{0},V}%
\]%
\begin{equation}
=-\frac{\partial}{\partial I_{0}}\frac{1}{2}\frac{d}{dr}<\left(
I^{(1)}\right)  ^{2}>_{\theta_{0},V}. \label{VTIVTT}%
\end{equation}

We assume that our perturbation is statistically uniform along the range $r$.
Then the right hand sides of Eqs. (\ref{I-1V}) and (\ref{VTIVTT}) does not
depend on range. Introducing the notation
\begin{equation}
B(I_{0})=\frac{d}{dr}<\left(  I^{(1)}\right)  ^{2}>_{\theta_{0},V},
\label{Bini}%
\end{equation}
combining Eqs. (\ref{dfidr}), (\ref{fiV}), (\ref{I-1V}), and (\ref{VTIVTT})
and replacing $I_{0}$ with $I$ we arrive at
\begin{equation}
\frac{d<\phi>}{dr}=\frac{1}{2}<B\frac{d^{2}\phi}{dI^{2}}>-\frac{1}{2}%
<\frac{dB}{dI}\frac{d\phi}{dI}>. \label{Fokker0}%
\end{equation}

Consider the probability density function (PDF) of action $W(I,r)$. Since
$\phi$ is an arbitrary function of $I$ and
\[
<\frac{d^{n}\phi}{dI^{n}}>=\int dI\,\phi(I)\,\frac{\partial^{n}}{\partial
I^{n}}W(I,r)
\]
for $n=0,1,...$ , Eq. (\ref{Fokker0}) implies that $W$ satisfies
\begin{equation}
\frac{\partial W}{\partial r}=\frac{1}{2}\frac{\partial}{\partial I}%
B\frac{\partial}{\partial I}W. \label{Fokker}%
\end{equation}
This is the Fokker-Planck equation for the PDF of the action. A more
convenient expression for the diffusivity $B$ may be obtained by substituting
Eq. (\ref{I-1}) into Eq. (\ref{Bini}). This yields
\begin{equation}
B=2\int_{r_{0}}^{r}dr^{\prime}<V_{\theta}^{(0)}(r)V_{\theta}^{(0)}(r^{\prime
})>_{V}. \label{B1}%
\end{equation}
Since $<V_{\theta}^{(0)}(r)V_{\theta}^{(0)}(r^{\prime})>$ becomes negligible
if $r-r^{\prime}\gg l_{m}$ and the above integral does not depend on the lower
limit $r_{0}$ provided $r-r_{0}\gg l_{m}$, we can formally replace $r_{0}$ in
Eq. (\ref{B1}) with $-\infty$.

Let us rewrite Eq. (\ref{B1}) in the form convenient for numerical evaluation
of $B$. Note that for the unperturbed trajectory with the action $I$
\begin{equation}
\frac{\partial z}{\partial\theta}=\frac{1}{\omega(I)}\frac{dz}{dr}=\frac
{1}{\omega(I)}\frac{p}{\sqrt{n^{2}-p^{2}}}\simeq\frac{p}{\omega(I)}.
\label{dzdT}%
\end{equation}
Exploit the approximation (\ref{V-delc}) and denote by $Z(I,r)$ and $P(I,r)$
the coordinate and momentum of an unperturbed ray trajectory with the action
$I$, respectively. Then Eq. (\ref{B1}) can be rewritten as
\begin{equation}
B(I)=\int_{-\infty}^{\infty}d\rho\,K_{\theta\theta}(I,\rho), \label{B}%
\end{equation}
where
\[
K_{\theta\theta}(I,\rho)=\frac{1}{R\omega^{2}(I)}%
\]%
\begin{equation}
\times\,\int_{0}^{R}dr\,\delta n_{z}(r)\delta n_{z}(r-\rho))P(I,r)P(I,r-\rho),
\label{KTT}%
\end{equation}
with
\begin{equation}
\delta n_{z}(r)=\left.  \frac{\partial\delta n(r,z)}{\partial z}\right|
_{z=Z(I,r)}. \label{deln-z}%
\end{equation}
It is assumed that the integral should be numerically calculated for a typical
realization of the inhomogeneity $\delta n$ over a range $R$ much greater than
both the correlation length of $\delta n(r,z)$ and the cycle length of the
unperturbed ray. The result should not depend (at least, should not depend
significantly) on the starting angle variable of the unperturbed ray and on a
particular realization of the inhomogeneity used in the numerical calculation.

\subsection{Statistical characteristics of perturbation\label{sec-stat}}

\paragraph{Evaluation of $<V>$.}

Consider the mean value $V$ at a fixed range $r$. Note that even if the
approximation (\ref{V-delc}) is valid, i.e. $V=-\delta n$, and $<\delta
n(r,z)>=0$ at any fixed point $(r,z)$ (we assume that this condition is met),
generally, $<V(I,\theta,r)>=-<\delta n(r,z(I,\theta)>\neq0$. The point is that
$I$ and $\theta$ at the range $r$ depend on all inhomogeneities at ranges
$r^{\prime}<r$, including those located within the interval $r-l_{m}%
<r^{\prime}<r.$ It means that $I$ and $\theta$ are, generally, correlated with
$\delta n$ at the range $r$ and when averaging $\delta n(r,z(I,\theta))$ we
cannot consider the argument $z(I,\theta)$ as a constant.

In order to find the mean perturbation $<V(I,\theta,r)>$ we again express $I$
and $\theta$ at the range $r$ through their values at the range $r_{0}$
satisfying the condition (\ref{lrl}) and will consider conditional averages
for a fixed value of $I_{0}$. Applying the expansion (\ref{I0-T0}) we find
that up to terms $O(V^{2})$
\[
<V(I,\theta,r)>_{V,\theta_{0}}=<V(I^{(0)},\theta^{(0)},r)>_{V,\theta_{0}}%
\]%
\begin{equation}
+<V_{I}^{(0)}I^{(1)}+V_{\theta}^{(0)}\theta^{(1)}>_{V,\theta_{0}}.
\label{Vmean0}%
\end{equation}
Since $I^{(0)}$ and $\theta^{(0)}$ are statistically independent of
inhomogeneities at the range $r$, the first term on the right of Eq.
(\ref{Vmean0}) vanishes. The second term may be transformed in the same manner
as we have transformed the terms on the right of Eq. (\ref{fiV}). Using Eqs.
(\ref{I-1})--(\ref{dV0dI0}) we get an analog to Eq. (\ref{av1})
\[
<V(I_{0},\theta,r)>_{V,\theta_{0}}=<I^{(1)}\frac{\partial}{\partial I_{0}%
}V^{(0)}%
\]%
\begin{equation}
+V_{\theta}^{(0)}\frac{\partial}{\partial I_{0}}\int_{r_{0}}^{r}dr^{\prime
}\,V^{(0)}(r^{\prime})>_{\theta_{0},V}. \label{av2}%
\end{equation}
By analogy with (\ref{uniform}) we have
\[
<\frac{\partial V^{(0)}(r)}{\partial\theta_{0}}\,V^{(0)}(r^{\prime}%
)>_{\theta_{0},V}%
\]%
\begin{equation}
=-<V^{(0)}(r)\,\frac{\partial V^{(0)}(r^{\prime})}{\partial\theta_{0}%
}>_{\theta_{0},V.} \label{uniform1}%
\end{equation}
Using this relation combined with Eq. (\ref{dTdT0}), Eq. (\ref{av2}) may be
transformed to
\[
<V(I_{0},\theta,r)>_{V,\theta_{0}}%
\]%
\begin{equation}
=\frac{\partial}{\partial I_{0}}\int_{-\infty}^{r}dr^{\prime}<V_{\theta}%
^{(0)}(r)V^{(0)}(r^{\prime})>_{\theta_{0},V}. \label{Vmin3}%
\end{equation}
A further averaging over $I_{0}$ provides $<V>$.

In the same manner as it has been done for the diffusivity (see the end of
Sec. \ref{sec-Fokker}) we present the expression for $<V> $ in the form
\begin{equation}
<V>=\int dI\, W(I,r)\, \frac{d}{dI}A_{1}(I), \label{meanV-1}%
\end{equation}
where
\begin{equation}
A_{1}(I)=\frac{1}{2}\int^{\infty}_{-\infty}d\rho\, K_{\theta}(I,\rho),
\label{meanV-2}%
\end{equation}
than for steep rays
\[
K_{\theta}(I,\rho)=\frac{1}{R\omega(I)}%
\]%
\begin{equation}
\times\, \int^{R}_{0}dr\, \delta n_{z}(r)\delta n(r-\rho)P(I,r). \label{KT}%
\end{equation}

\paragraph{Evaluation of $<V_{I}>$.}

In a similar way it can be shown that
\begin{equation}
<V_{I}>=\int dI\, W(I,r)\, \frac{d}{dI}A_{2}(I), \label{meanVI-1}%
\end{equation}
where
\begin{equation}
A_{2}(I)=\frac{1}{2}\int^{\infty}_{-\infty}d\rho\, K_{\theta I}(I,\rho),
\label{meanVI-2}%
\end{equation}%
\[
K_{\theta I}(I,\rho)=\frac{1}{R\omega(I)}%
\]%
\begin{equation}
\times\, \int^{R}_{0}dr\, \delta n_{z}(r)\delta n_{z}(r-\rho))P(I,r-\rho
)Z_{I}(I,r), \label{KTI}%
\end{equation}%
\[
Z_{I}(I,r)=\frac{\partial}{\partial I}Z(I,r).
\]

\paragraph{Evaluation of $<\left(  \int Vdr\right)  ^{2}>$.}

It is clear that the mean integral
\begin{equation}
<\int_{0}^{r}V(I,\theta,r)\,dr^{\prime}>=\int_{0}^{r}<V(I,\theta
,r)>\,dr^{\prime} \label{int-alpha}%
\end{equation}
may be estimated using Eqs. (\ref{meanV-1})-(\ref{KTT}). By analogy with the
expressions for the diffusivity and $<V>$ derived above we easily obtain
\begin{equation}
<\left(  \int_{0}^{r}V(I,\theta,r)\,dr^{\prime}\right)  ^{2}>=r\,\int
_{-\infty}^{\infty}d\rho\,K(I,\rho), \label{V2mean}%
\end{equation}
where
\begin{equation}
K(I,\rho)=\frac{1}{R}\int_{0}^{R}dr\,\delta n(r)\delta n(r-\rho)). \label{K}%
\end{equation}

\paragraph{Evaluation of $<\left(  \int V_{I}dr\right)  ^{2}>$.}

Similarly,
\begin{equation}
<\left(  \int^{r}_{0}V_{I}(I,\theta,r)\, dr^{\prime}\right)  ^{2}>=r\,
\int^{\infty}_{-\infty}d\rho\, K_{II}(I,\rho), \label{V-I2mean}%
\end{equation}
where
\[
K_{II}(I,\rho)=\frac{1}{R}%
\]%
\begin{equation}
\times\, \int^{R}_{0}dr\, \delta n_{z}(r)\delta n_{z}(r-\rho))Z_{I}%
(r)Z_{I}(r-\rho). \label{KII}%
\end{equation}

\subsection{Approximation of the action variable by a Wiener process}

\begin{figure}[ptb]
{}
\par
\centering\includegraphics[width=7cm,keepaspectratio=true]{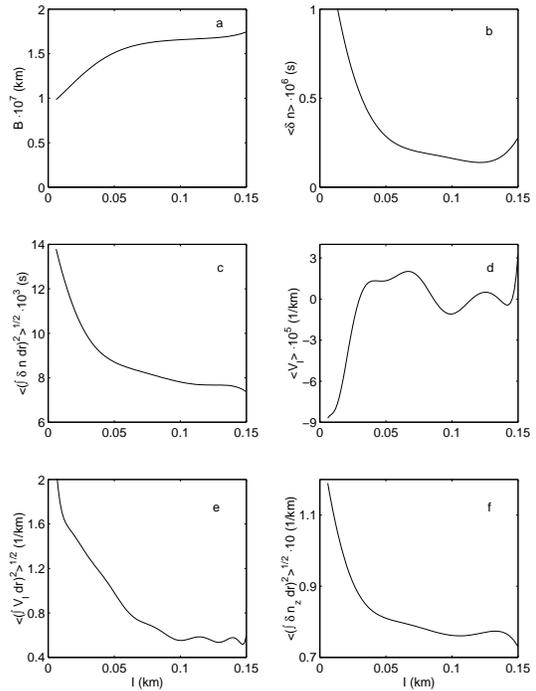}%
\caption{Statistical characteristics numerically evaluated over the
unperturbed ray trajectory versus the action $I$. The integrals presented in
panels c, e and f are evaluated over the 3000 km range.}%
\label{statist}%
\end{figure}Figures 6 a-f present some statistical characteristics numerically
evaluated using equations derived in Secs. \ref{sec-Fokker} and \ref{sec-stat}%
. All the characteristics have been calculated by integrating over unperturbed
trajectories placed into realizations of the perturbed waveguide. Consistent
with our expectation results only slightly depend on particular realizations
of the inhomogeneity used in calculations. The characteristics are considered
as functions of the action variable determining the shape of the unperturbed trajectory.

In Fig. 6a it is seen that the dependence of the diffusivity $B$ on $I$ is
comparatively weak and we can approximate it by a constant
\begin{equation}
B=1.5\,10^{-7}\text{ km.} \label{Bconst}%
\end{equation}
Due to the smallness of perturbation the term on the right of Eq.
(\ref{dIdrH}) varies on scales small to that of the range variations of $I$
and in this sense the term $V_{\theta}$ can be considered as delta-correlated.
Therefore we can present the action as
\begin{equation}
I(r)=I_{s}+x(r), \label{Ipert}%
\end{equation}
where $I_{s}=I(0)$ and $x(r)$ is governed by equation
\begin{equation}
\frac{dx}{dr}=\xi(r), \label{dxdr1}%
\end{equation}
where $\xi$ is a white noise with
\begin{equation}
<\xi>=0,\;\;\;<\xi(r)\xi(r^{\prime})>=B\delta(r-r^{\prime}). \label{xi_cor}%
\end{equation}
Here $B$ is a constant defined by Eq. (\ref{Bconst}). Equations (\ref{Ipert}%
)-(\ref{xi_cor}) idealize a random increment of the action, $x(r)$, as a
Wiener process \cite{Tikhonov77} with%

\begin{equation}
<x>=0,\;<x(r_{1})\,x(r_{2})>=B_{\xi}(I)\,\min(r_{1},r_{2}). \label{x_corr}%
\end{equation}
The variance of $x$ is a linear function of range%

\begin{equation}
\sigma_{I}^{2}\equiv<x^{2}>=B\,r. \label{sig_x}%
\end{equation}

Turn our attention to the angle variable. Representing $\theta$ as
\begin{equation}
\theta(r)=\theta_{s}+\omega(I_{s})r+y(r), \label{Tpert}%
\end{equation}
substituting this and Eq. (\ref{Ipert}) into Eq. (\ref{dtedrH}) we get
\begin{equation}
\frac{dy}{dr}=\omega(I_{s}+x)-\omega(I_{s})+V_{I}. \label{dydr0}%
\end{equation}
Simplify this equation using the following two small parameters in the problem.

\textbf{(i) Parameter }$\mu$. It is defined as
\begin{equation}
\mu=\frac{\sigma_{I}}{I^{\ast}}=\left(  \frac{\omega^{\prime}}{\omega}\right)
_{\text{rms}}\sqrt{Br}, \label{mu}%
\end{equation}
with $I^{\ast}=\left|  \omega^{\prime}/\omega\right|  _{\text{rms}}^{-1}$
being a characteristic scale of the smooth function $\omega(I)$. The smallness
of $\mu$ allows one to replace $\omega(I_{s}+x)-\omega(I_{s})$ with
$\omega^{\prime}(I_{s})x$. It is important to emphasize that, typically,
$I_{s}\ll I^{\ast}$ and the condition
\begin{equation}
\mu\ll1 \label{mull1}%
\end{equation}
does not imply that the magnitude of action fluctuations is small compared to
$I_{s}$. In our model typical values of $\mu$ at a 3000 km range are $0.1 $
for flat rays and $0.01$ for steep ones.

\textbf{(ii) Small parameter }$\bar{\mu}$. This parameter estimates relative
contributions to $y$ from two terms on the right of Eq. (\ref{dydr0}):
\[
\bar{\mu}=\frac{<\left(  \int_{0}^{r}V_{I}dr^{\prime}\right)  ^{2}>^{1/2}%
}{\left|  \omega^{\prime}\right|  <\left(  \int_{0}^{r}x(r^{\prime}%
)dr^{\prime}\right)  ^{2}>^{1/2}}%
\]%
\begin{equation}
=\frac{<\left(  \int_{0}^{r}V_{I}dr^{\prime}\right)  ^{2}>^{1/2}}{\left|
\omega^{\prime}\right|  \,\left(  Br^{3}/3\right)  ^{1/2}}. \label{mubar}%
\end{equation}
Values of $<\left(  \int_{0}^{r}V_{I}dr^{\prime}\right)  ^{2}>^{1/2}$ at a
3000 km range are shown in Fig. 6e.

Note that according to (\ref{V-delc}) $V_{I}=-\delta n_{z}\,z_{I}$, where
$\delta n_{z}\equiv\partial\delta n/\partial z$, $z_{I}\equiv\partial
z/\partial I$. An order-of-magnitude estimate of $z_{I}$ can be obtained in
the following way. At $\theta=\pi$ and $\theta=2\pi$ the unperturbed ray has
turning points. For these values of $\theta$ we have $n(z)=-H$, and
\begin{equation}
n_{z}z_{I}=-\omega, \label{zI}%
\end{equation}
where $n_{z}\equiv\partial n/\partial z$. Denote order-of-magnitude estimates
of $z_{I}$, $n_{z}$,$\delta n$ and $\delta n_{z}$ by $z_{I}^{\ast}$,
$n_{z}^{\ast}$, $\delta n^{\ast}$ and $\delta n_{z}^{\ast}$, respectively. In
our model%
\[
n_{z}^{\ast}=0.015\text{ 1/km, \ \ }\delta n^{\ast}=5\cdot10^{-7},
\]
\begin{equation}
\delta n_{z}^{\ast}=0.08\,\,\,1/\text{km}. \label{dndnz}%
\end{equation}
Since a typical value of $\omega$ is about $0.15$ 1/km, Eqs. (\ref{zI}) and
(\ref{dndnz}) give $z_{I}^{\ast}=10$. A rough estimate of $<\left(  \int
_{0}^{r}V_{I}dr^{\prime}\right)  ^{2}>^{1/2}$is given by $<\left(  \int
_{0}^{r}\delta n_{z}\,dr^{\prime}\right)  \,^{2}>^{1/2}$ times $z_{I}^{\ast}$.
Using data presented in Figs. 6e and 6f it follows that both this rough
estimate and the direct evaluation of $<\left(  \int_{0}^{r}V_{I}dr^{\prime
}\right)  ^{2}>^{1/2}$give approximately the same result. Taking the value of
$B$ from Eq. (\ref{Bconst}) and values of $\omega^{\prime}$ from Fig. 15
presented below we find that at a 3000 km range the parameter $\bar{\mu}$
varies within the interval of $0.1$ for steep rays to $0.01$ for flat ones.

Due to smallness of $\mu$ and $\bar{\mu}$ we can simplify Eq. (\ref{dydr0}) by
neglecting the last term on the right and retaining only the first order in
the expansion of $\omega(I_{s}+x)-\omega(I_{s})$. This yields
\begin{equation}
\frac{dy}{dr}=\omega^{\prime}(I_{s})x, \label{dydr}%
\end{equation}
i.e. the random component of the angle variable is an integral of a Wiener
process
\begin{equation}
y(r)=\omega^{\prime}(I_{s})\,\int_{0}^{r}x(r_{1})dr_{1}. \label{y_via_x}%
\end{equation}
Its variance
\begin{equation}
\sigma_{\theta}^{2}\equiv<y^{2}>=(\omega^{\prime})^{2}\,B\frac{r^{3}}{3}.
\label{sig_y}%
\end{equation}

Thus, we have replaced exact ray equations (\ref{dIdrH}) and (\ref{dtedrH}) by
remarkably simple approximate stochastic equations (\ref{dxdr1}) and
(\ref{dydr}).

To demonstrate accuracy of Eqs. (\ref{sig_x}) and (\ref{sig_y}) consider a
numerical example. Figure 7 graphs standard deviations of action and angle
variables, i.e. $\sigma_{I}$ and $\sigma_{\theta}$, respectively, for a ray
path starting at a depth of $0.78$ km. Solid curves present results of
averaging over a fan of 100 rays with launch angles from a narrow interval
centered at $\chi_{c}=7.8^{\circ}$ (corresponding action $I_{c}=$ $0.06$ km).
Dashed curves show dependencies given by Eqs. (\ref{sig_x}) and (\ref{sig_y}).
It is seen that the simple statistical model considered in this section
provides reasonable predictions for standard deviations of $x$ and $y$.
Similar results have been obtained for rays starting at different launch angles.

Relations (\ref{sig_x}) and (\ref{sig_y}) describe statistics of a cluster of
rays with launch angles close to some fixed value. This result will be used
below in Sec. \ref{structure}. \begin{figure}[ptb]
\begin{center}
\includegraphics[width=7.5cm,height=6.4cm]{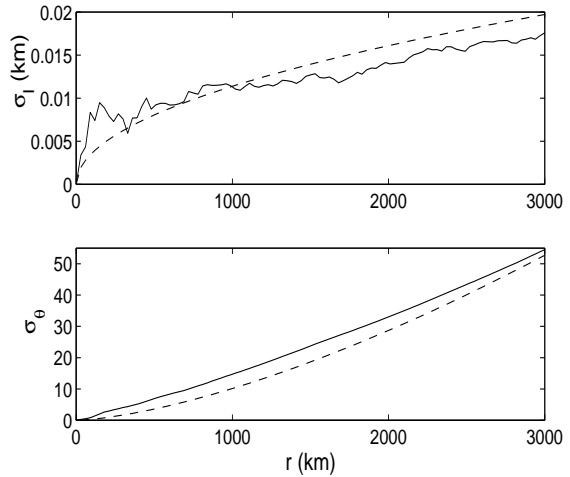}
\end{center}
\caption{Standard deviations of the action (upper panel) and angle (lower
panel) variables ($\sigma_{I}$ and $\sigma_{\theta}$, respectively) as
functions of range computed for a fan of 100 rays escaping a point source set
at a depth of $0.78$ km. Launch angles of the rays span a narrow angular
interval centered at $7.8^{\circ}$. The solid curves present results of
averaging over the fan rays. The estimation $\sigma_{I}$ given by Eq.
(\ref{sig_x}) is used to produce the dashed line in the upper panel. The
dashed line in the lower panel is the estimation for $\sigma_{\theta}$ given
by Eq. (\ref{sig_y}).}%
\label{sigma}%
\end{figure}

\subsection{Cluster of rays with a given identifier\label{cluster}}

Now let us focus on a cluster of different type. Consider rays leaving a point
source and arriving at the given range $r$ with the given identifier. \bigskip
Clusters of this type with identifiers $+124$ and $+160$ form two groups of
arrivals shown by thick points in Figs. 2 and 3, respectively.

To obtain an approximate analytical description of such clusters, compare two
rays -- one in the perturbed waveguide and another in the unperturbed
waveguide -- with close but, generally, different starting action variables.
Using Eqs. (\ref{Tpert}) and (\ref{y_via_x}), we present angle variables of
these rays at the range $r$ as
\[
\theta=\theta_{s}+\omega(I_{s})r+\omega^{\prime}(I_{s})\int_{0}^{r}%
x(r_{1})dr_{1}%
\]
for the ray in the perturbed waveguide and
\[
\bar{\theta}=\bar{\theta}_{s}+\omega(\bar{I})r
\]
for the ray in the unperturbed waveguide. The difference $\theta-\bar{\theta}$
can be approximated by the relation
\[
\theta-\bar{\theta}\simeq\int_{0}^{r}\left[  \omega(I_{s}+x(r_{1}%
))-\omega(\bar{I})\right]  dr_{1}%
\]%
\begin{equation}
\simeq\bar{\omega}^{\prime}(I_{s})\left(  (I_{s}-\bar{I})r+\int_{0}^{r}%
x(r_{1})dr_{1}\right)  , \label{T_Tbar}%
\end{equation}
where we have omitted the constant $\theta_{s}-\bar{\theta}_{s}$. At long
ranges this constant becomes negligible compared to other terms whose rms
magnitudes grow, on average, with $r$. Then the equation
\begin{equation}
\;\;\frac{1}{r}\int_{0}^{r}x(r_{1})dr_{1}=-I_{s}+\bar{I} \label{q}%
\end{equation}
can be idealized as a condition that singles out perturbed rays whose
identifiers at the range $r$ are equal to the identifier of the unperturbed
ray with the action $\bar{I}$. Using this condition combined with Eqs.
(\ref{Ipert}), (\ref{dxdr1}) and (\ref{xi_cor}), one can evaluate statistical
characteristics of action variables of rays belonging to the cluster. To
illustrate this statement we consider a couple of examples.

First, evaluate the probability density function (PDF) $P(I_{s})$, defined in
a following way: $P(I_{s})dI_{s}$ is a probability that a stochastic ray with
a starting action within the interval of $I_{s}$ to $I_{s}+dI_{s}$ meets the
condition (\ref{q}), i.e. this ray belongs to the cluster defined by Eq.
(\ref{q}). In other words, we consider $I_{s}$ as a random variable and our
task is to find its PDF.

It is well-known that the PDF of a random variable $\beta$ can be presented in
the form
\begin{equation}
P(b)=<\delta(b-\beta)> \label{P_delta}%
\end{equation}
where the angular brackets denote the ensemble averaging operation. Note, that
we denote any PDF by the same symbol $P$ with an argument indicating a
specific random variable.

Making use of Eq. (\ref{P_delta}) yields
\begin{equation}
P(I_{s})=<\delta\left(  I_{s}-\bar{I}+\frac{1}{r}\int_{0}^{r}x(r_{1}%
)dr_{1}\right)  >, \label{delta_aver}%
\end{equation}
where the averaging goes over trajectories of the Wiener process $x(r)$
defined by Eqs. (\ref{dxdr1}) and (\ref{xi_cor}) with the initial condition
$x(0)=0$.

Using the Fourier-representation of the $\delta$-function we rewrite Eq.
(\ref{delta_aver}) as
\begin{equation}
P(I_{s})=\frac{1}{2\pi}\int_{-\infty}^{\infty}d\gamma e^{i\gamma(I_{s}-\bar
{I})}<e^{i\frac{\gamma}{r}\int_{0}^{r}x(r_{1})dr_{1}}>. \label{P_Is_1}%
\end{equation}
The integral
\begin{equation}
g=\frac{1}{r}\int_{0}^{r}x(r_{1})dr_{1} \label{g}%
\end{equation}
is a Gaussian random variable with zero mean, $<g>=0$. The average in the
integrand on the right of Eq. (\ref{P_Is_1}) can be rewritten as
\cite{Rytov76}
\begin{equation}
<e^{i\gamma g}>=e^{-\frac{\gamma^{2}}{2}<g^{2}>}. \label{Gauss_aver}%
\end{equation}
Substituting Eq. (\ref{Gauss_aver}) in Eq. (\ref{P_Is_1}) we arrive at a
Gaussian integral over $\gamma$. Evaluating this integral and applying Eq.
(\ref{x_corr}) to find $<g^{2}>$ yields
\begin{equation}
P(I_{s})=\sqrt{\frac{3}{2\pi Br}}\exp\left[  -\frac{3}{2}\frac{(I_{s}-\bar
{I})^{2}}{Br}\right]  . \label{P_Is}%
\end{equation}
The standard deviation of $I_{s}$ from its mean value $\bar{I}$ is given by
\begin{equation}
\sigma_{I_{s}}=\sqrt{Br/3}. \label{sig_Is}%
\end{equation}

In the scope of our statistical model the difference in actions $\Delta
I(\rho)=I-\bar{I}$ $\ $present in equations for $\Delta S$ and $\Delta t$
derived in Sec. \ref{travel}, is a Gaussian random function with a zero mean
whose statistical characteristics are completely determined by the correlation
function \cite{Rytov76}
\begin{equation}
Q(\rho_{1},\rho_{2})=<\Delta I(\rho_{1})\,\Delta I(\rho_{2})>. \label{Q_def}%
\end{equation}
Note, that $\Delta I(\rho)=$ $I_{s}+x(\rho)-\bar{I}$. For short, we shall use
the notation $a_{1,2}=\Delta I(\rho_{1,2})$. Applying again Eq. (\ref{P_delta}%
) we obtain the following expression for the joint PDF of $I_{s}$, $a_{1}$,
and $a_{2}$ (it is assumed that $0\leq\rho_{1,2}\leq r$):
\[
P(I_{s},a_{1},a_{2})=<\delta\left(  I_{s}-\bar{I}+g\right)
\]%
\begin{equation}
\times\delta(a_{1}-I_{s}+\bar{I}-x(\rho_{1}))\,\,\delta(a_{2}-I_{s}+\bar
{I}-x(\rho_{2}))>. \label{P_Is_a1_a2}%
\end{equation}
Then
\[
Q(\rho_{1},\rho_{2})=\int\,a_{1}a_{2}P(I_{s},a_{1},a_{2})\,dI_{s}da_{1}da_{2}%
\]%
\[
=<(x(\rho_{1})-g)(x(\rho_{2})-g)>
\]%
\begin{equation}
=B\left(  \frac{r}{3}-\rho_{1}-\rho_{2}+\frac{\rho_{1}^{2}+\rho_{2}^{2}}%
{2r}+\min(\rho_{1},\rho_{2})\right)  . \label{Q_explicit}%
\end{equation}

Later on we shall use this equation to calculate statistical moments of
$\Delta I$ in order to estimate individual terms in the expansion
(\ref{Del_t_I}). In so doing we shall apply the known properties of
statistical momenta of Gaussian random variables \cite{Rytov76}. In
particular, we shall use the relation
\[
<\Delta I^{2}(\rho_{1})\Delta I^{2}(\rho_{2})>=<\Delta I^{2}(\rho_{1})><\Delta
I^{2}(\rho_{2})>
\]%
\[
+2<\Delta I(\rho_{1})\Delta I(\rho_{2})>^{2}%
\]%
\begin{equation}
=Q(\rho_{1},\rho_{1})Q(\rho_{2},\rho_{2})+2Q^{2}(\rho_{1},\rho_{2}).
\label{M4}%
\end{equation}

\subsection{Stochastic instability of ray identifier}

In Sec. \ref{timefront} we have already seen that in the perturbed waveguide
ray trajectories exhibit extreme sensitivity to starting launch angles. An
impressive demonstration of this instability is presented in Fig. 5. The
points representing dependence of the ray identifier on the launch angle are
randomly scattered and this fact suggests that rays with very close launch
angles are practically uncorrelated and should be described statistically.

Our stochastic ray theory provides a tool for quantitative description of
stochastic ray instability. In particular, it allows one to estimate chaotic
spread of ray identifiers shown in Fig. 5. Consider unperturbed and perturbed
rays starting at the same launch angle, $\chi_{s}$, and, consequently, with
the same starting action, $I_{s}$. Denoting the numbers of turning points of
these paths by $\bar{J}$ and $J$, respectively, we note that at long ranges
both $\bar{J}$ and $J$ are large and the same is true of their rms difference.
Then we can approximate $\Delta J=J-\bar{J}$ by $y/2\pi$. This yields $<\Delta
J>=0$ and $\sigma_{\Delta J}\equiv(<\Delta J^{2}>)^{1/2}=\sigma_{\theta}/2\pi$
with $\sigma_{\theta}$ given by Eq. (\ref{sig_y}). Assuming that for most rays
with starting angles close to $\chi_{s}$ the value of $J$ lies within the
interval
\begin{equation}
\bar{J}-2\sigma_{\Delta J}<J<\bar{J}+2\sigma_{\Delta J}, \label{Jspread}%
\end{equation}
we find an estimate for the spread of ray identifiers. Figure 8 presents the
number of ray turning points, $J$, against the launch angle, $\chi_{s}$, at
1500 km and 3000 km. For 3000 km range we have the same plot as in Fig. 5, but
the solid line representing the dependence of $\bar{J}$ on $\chi_{s}$ is
slightly smoothed. The dashed lines represent limits defined by Eq.
(\ref{Jspread}). This result confirms that Eq. (\ref{Jspread}) gives a
reasonable estimation of the spread of ray identifiers.

Stochastic dependence of the identifier on the launch angle can be considered
from a different viewpoint. Let us fix some ray identifier and study
statistics of starting actions, $I_{s}$, of perturbed rays arriving at the
given range $r$ with this identifier. The probability density function of
$I_{s}$ is given by Eq. (\ref{P_Is}). The mean value of $I_{s}$ is equal to
$\bar{I}$, action of an unperturbed ray which has the given identifier at the
range $r$. According to Eq. (\ref{P_Is}) $I_{s}$ is a Gaussian random variable
and it is natural to expect that most part of rays with the given identifier
at range $r$ have starting action within the interval
\begin{equation}
\bar{I}-2\sigma_{I_{s}}<I_{s}<\bar{I}+2\sigma_{I_{s}}, \label{Is_spread}%
\end{equation}
where $\sigma_{I_{s}}$ is determined by Eq. (\ref{sig_Is}). Figure 9
demonstrates that this prediction agrees with results of our numerical
simulation. The solid lines present dependencies of the starting action of the
unperturbed ray on the number of ray turning points at 1500 km and 3000 km,
while the dashed lines indicate the borders of intervals defined by Eq.
(\ref{Is_spread}). Consistent with our expectation, most points depicting
parameters $I_{s}$ of perturbed rays starting upward against numbers of their
turning points at 1500 km (upper panel) and 3000 km (lower panel) lie within
areas embraced by the dashed lines. \begin{figure}[ptb]
\begin{center}
\includegraphics[width=7cm,keepaspectratio=true]{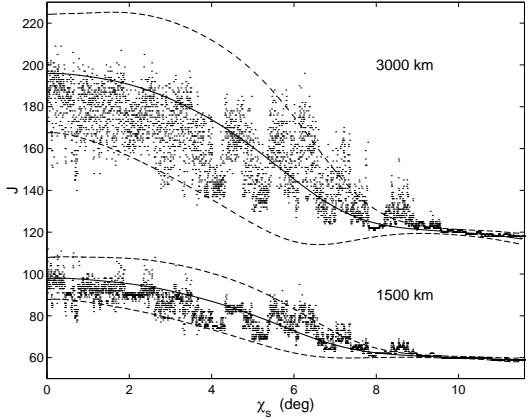}
\end{center}
\par
.\caption{Ray identifier, $J$, as a function of the launch angle, $\chi_{s}$,
at the ranges 1500 and 3000 km. The points depicting rays at 3000 km range are
the same as in Figs. 2 and 3. The solid lines are smoothed functions
$J(\chi_{s})$ for the unperturbed waveguide. The dashed lines show the limits
established by Eq. (\ref{Jspread})}%
\label{JVSchi}%
\end{figure}\begin{figure}[ptbptb]
\begin{center}
\includegraphics[width=7cm,keepaspectratio=true]{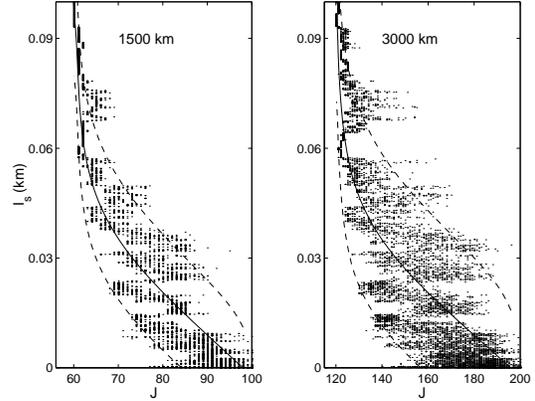}
\end{center}
\caption{Starting action, $I_{s}$, as a function of ray identifier, $J$, for
the ranges 1500 km (left) and 3000 km (right). The solid lines are smoothed
functions $I_{s}(J)$ for the unperturbed waveguide. The dashed lines show the
limits established by Eq. (\ref{Is_spread}).}%
\label{IvsJ}%
\end{figure}

\section{Small parameters in the problem and approximate formula for travel
time variations\label{small_parameters}}

The approximate stochastic ray theory developed in the preceding section
allows one to deduce simple analytical estimations of terms present on the
right side of Eq. (\ref{constituents}). Our main concern is with terms $\Delta
t_{I}$ and $\Delta t_{V}$ which are given by integrals over $r$ and, in
principle, may be very large at long ranges. The objective of the present
section is to compare these constituents of $\Delta t$ and simplify Eq.
(\ref{constituents}) by neglecting small terms.

\textbf{(i) Estimation of }$\mathbf{\Delta t}_{I}$\textbf{. }Divide $\Delta
t_{I}$ into a sum
\begin{equation}
\Delta t_{I}=\Delta t_{I}^{(2)}+\Delta t_{I}^{(3)}+\Delta t_{I}^{(4)}+\ldots,
\label{Del_tI_split}%
\end{equation}
where
\begin{equation}
\Delta t_{I}^{(2)}=\,\frac{\bar{\omega}^{\prime}}{2c_{r}}\int\Delta
I^{2}\,\,dr, \label{Del_tI_2}%
\end{equation}%
\begin{equation}
\Delta t_{I}^{(3)}=\,\frac{\bar{\omega}^{\prime\prime}}{3c_{r}}\int\,\Delta
I^{3}\,dr, \label{Del_tI_3}%
\end{equation}%
\begin{equation}
\Delta t_{I}^{(4)}=\,\frac{\bar{\omega}^{\prime\prime\prime}}{8c_{r}}%
\int\Delta I^{4}\,dr, \label{Del_tI_4}%
\end{equation}
and estimate mean values of components $\Delta t_{I}^{(2)}$, $\Delta
t_{I}^{(3)}$, and $\Delta t_{I}^{(4)}$. Applying Eq. (\ref{Q_explicit}) to
find $<\Delta I^{2}>$ and making use of the relation
\begin{equation}
<\Delta I^{4}>=3<\Delta I^{2}>^{2} \label{I4_Gauss}%
\end{equation}
valid for a Gaussian random variable with a zero mean (it follows from Eq.
(\ref{M4})), yields
\begin{equation}
<\Delta t_{I}^{(2)}>=\frac{\bar{\omega}^{\prime}}{12c_{r}}Br^{2},
\label{Del_tI_aver}%
\end{equation}%
\begin{equation}
\;\;<\Delta t_{I}^{(3)}>=0,\;\;<\Delta t_{I}^{(4)}>=\,\frac{\bar{\omega
}^{\prime\prime\prime}}{80\,c_{r}}B^{2}r^{3}. \label{Del_tI_aver34}%
\end{equation}

\textbf{(ii) Estimation of }$\mathbf{\Delta t}_{V}$\textbf{. }Due to smallness
of $\Delta I$ we can transform the integrand on the right of Eq.
(\ref{Del_t_V}) as
\begin{equation}
\Delta I\,V_{I}(I,\theta,r)-V\simeq-V(\bar{I},\theta,r)=\delta n(r,z(\bar
{I},\theta)). \label{tV-integrand}%
\end{equation}
Then
\begin{equation}
c_{r}\Delta t_{V}=\int\delta n(r,z(\bar{I},\theta))\,dr. \label{Del-tVm}%
\end{equation}
The integration in Eq. (\ref{Del-tVm}) goes along the unperturbed ray path
determined by the action $\bar{I}$. Since $\theta$, generally, differs from
$\bar{\theta}$, this integral differs from that present in Eq. (\ref{V2mean})
which is evaluated over the unperturbed ray trajectory. However it is clear,
that the statistics of $\Delta t_{V}$ determined by Eq. (\ref{Del-tVm}) is
approximately described by Eqs. (\ref{V2mean}), and (\ref{K}). Figure Fig. 6c
shows the standard deviation of $\Delta t_{V}/c_{r}$ at the 3000 km range.
Note that the quantity similar to $\Delta t_{V}$ evaluated over an unperturbed
ray path, is widely used to estimate travel time shifts due to inhomogeneities
at short enough ranges \cite{FDMWZ79}.

Using Eqs. (\ref{Del_tI_aver}) and (\ref{Del_tI_aver34}) we see that the first
term in the sum (\ref{Del_tI_split}) dominates if the parameter%

\begin{equation}
\mu_{1}=\left|  \frac{<\Delta t_{I}^{(4)}>}{<\Delta t_{I}^{(2)}>}\right|
=\frac{3}{20}Br\left|  \frac{\bar{\omega}^{\prime\prime\prime}}{\ \bar{\omega
}^{\prime}}\right|  \label{mu1}%
\end{equation}
is small compared to unity. Taking into account Eq. (\ref{sig_x}) we can
roughly estimate $\mu_{1}$ as
\begin{equation}
\mu_{1}\approx\frac{\sigma_{I}^{2}}{I^{\ast\ast2}}, \label{mu1_rough}%
\end{equation}
where $I^{\ast\ast}$ is some characteristic scale. It turns out that for our
environmental model $I^{\ast\ast}$ differs considerably from the scale
$I^{\ast}$ present in Eq. (\ref{mu}). Therefore, the parameter $\mu_{1}$
cannot be estimated as $\mu^{2}$ as it might be expected. Nevertheless,
smallness of both $\mu$ and $\mu_{1}$ is caused by the same factors: the
weakness of the perturbation combined with the smoothness of the function
$\omega(I)$.

The smallness of $\mu_{1}$ allows one to replace the constituent of the
difference in ray travel times $\Delta t_{I}$ (see Eqs. (\ref{constituents})
and (\ref{Del_t_I})) with $\Delta t_{I}^{(2)}$. Taking into account
(\ref{Del-tVm}) we can now represent the general expression for the difference
in travel time (\ref{constituents}) in the form
\begin{equation}
c_{r}\Delta t=2\pi\,\Delta N\,\bar{I}+c_{r}\Delta t_{G}+\frac{\omega^{\prime}%
}{2}\int\Delta I^{2}dr+\int\delta n\,dr. \label{Del_t_short}%
\end{equation}
This formula is the main result of the present work. Note that it can be
further simplified if we consider two eigenrays coinciding the same pair of
the endpoints in the perturbed and unperturbed waveguides. In this case
$z_{s}=\bar{z}_{s}$ and $z_{e}=\bar{z}_{e}$ and according Eq. (\ref{Del_t_G})
\begin{equation}
\Delta t_{G}=0. \label{Del-t-G1}%
\end{equation}
If both rays leave the same point, $z_{s}=\bar{z}_{s}$, and arrive at two
different points whose depths are close, then using Eq. (\ref{canon1H}) we can
approximately present $\Delta t_{G}$ as
\begin{equation}
\Delta t_{G}=\bar{p}_{e}\Delta z_{e}. \label{Del-t-G2}%
\end{equation}

If parameter
\begin{equation}
\mu_{2}=\frac{<(\Delta t_{V})^{2}>^{1/2}}{\left|  <\Delta t_{I}^{(2)}>\right|
}=\frac{12c_{r}<(\Delta t_{V})^{2}>^{1/2}}{\left|  \omega^{\prime}\right|
Br^{2}} \label{mu2}%
\end{equation}
is small compared to unity, then the last term on the right of Eq.
(\ref{Del_t_short}), $\Delta t_{V}$, can be neglected because it is small
compared to $\Delta t_{I}$.

Figure 10 shows parameters $\mu_{1}$and $\mu_{2}$ at the range 3000 km as
functions of the launch angle for a point source set at a depth of $0.78$ km.
It is seen that while $\mu_{1}$ is everywhere small, the parameter $\mu_{2}$
is small only for flat enough rays.

\begin{figure}[ptb]
\begin{center}
\includegraphics[width=7cm,keepaspectratio=true]{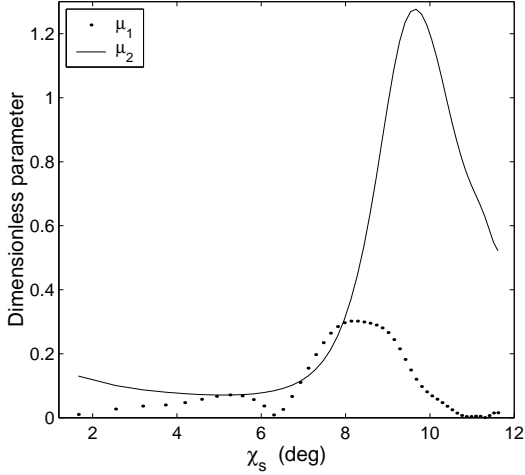}
\end{center}
\caption{Dimensionless parameters defined by Eqs. (\ref{mu1}) and (\ref{mu2})
as functions of launch angle at the 3000 km range.}%
\label{mus}%
\end{figure}

In what follows we shall apply Eq. (\ref{Del_t_short}) to study variations of
the timefront due to internal-wave-induced random inhomogeneities. Our primary
concern will be with the estimation of widening and bias of the timefront
segments in the presence of perturbation.

\section{Analytical description of timefront structure\label{structure}}

\subsection{\bigskip Timefront in a range-independent
waveguide\label{range_indep}}

Although Eq. (\ref{Del_t_short}) has been derived to compare travel times of
perturbed and unperturbed rays, it can also be applied in the case when both
rays propagate in the unperturbed waveguide with $\delta n=0$. Then the fourth
term on the right in Eq. (\ref{Del_t_short}) vanishes, the action $I$ (like
$\bar{I}$) does not depend on $r$, and the same is true of $\Delta I$. We
shall compare rays arriving at the same points, i.e. eigenrays, and in
accordance with Eq. (\ref{Del_t_G}) neglect the term $\Delta t_{G}$. This
yields
\begin{equation}
c_{r}\Delta t=2\pi\bar{I}\,\Delta N+\frac{\bar{\omega}^{\prime}}{2}\Delta
I^{2}r\text{.}\label{Del_t_hom}%
\end{equation}
In the case $\Delta N=N-\bar{N}=1$ this equation provides a difference in
travel times of two eigenrays with identifiers $\pm J$ and $\pm(J-2)$.
According to Eqs. (\ref{I0H}) and (\ref{theta}) the action variable of these
eigenrays satisfies
\[
\omega(I)-\omega(\bar{I})=\bar{\omega}^{\prime}\,\Delta I+O(\Delta I^{2})
\]%
\begin{equation}
=\frac{1}{r}(2\pi-\Delta\theta_{s}+\Delta\theta_{e}).
\end{equation}
Assuming that $\omega$ as a function of $I$ is monotonous at the interval
$(I,\bar{I})$, at long ranges ($N\gg1$) we have $\left|  \Delta\theta
_{s}\right|  ,\,\left|  \Delta\theta_{e}\right|  \ll2\pi$ and
\begin{equation}
\Delta I=\frac{2\pi}{\bar{\omega}^{\prime}r}+O\left(  \frac{1}{r^{2}}\right)
.
\end{equation}
Then Eq. (\ref{Del_t_hom}) reduces to
\begin{equation}
c_{r}\Delta t=\pi(I+\bar{I}).\label{I_Ibar}%
\end{equation}

A similar result have been obtained in Refs. \cite{V85,MW83,V95} (in Refs.
\cite{V85,V95} it has been derived for an adiabatically range-dependent waveguide).

An interesting and somewhat surprising fact following from Eq. (\ref{I_Ibar})
is that there exists a conservation law for temporal shifts between timefront
segments. Consider a bunch of rays with launch angles within a narrow
interval. Action variables of all these rays are close to some value which we
denote by $I_{0}$. Beginning from a certain range $r_{\ast}$ these rays will
form at least two segments with identifiers that differ by 2. When estimating
the temporal shift between two such segments we shall compare rays arriving at
the same depths. With this in mind, we can estimate the temporal shift as
$\tau_{0}=2\pi I_{0}/c_{r}$. It should be emphasized that $\tau_{0}$ does
\textbf{not} depend on range. It means that although the number of segments
formed by rays with launch angles from a given narrow angular interval grows
linearly with range, temporal shifts between neighboring segments (to be more
precise, between segments corresponding to identifiers $\pm J$ and $\pm(J-2)$
with $J$ depending on range ) remain approximately the same at any distance. A
more detailed discussion of this issue is given in Refs. \cite{V85,V95}.

The above statement means that a simple evaluation of the action variable as a
function of launch angle, $I(\chi)$, gives a considerable quantitative
information on temporal structure of the pulse signal valid at arbitrary (long
enough) range. If $N\gg1$, the values of $I\,$\ and $\bar{I}$ on the right
side of Eq. (\ref{I_Ibar}) are close and this equation can be approximated by
\begin{equation}
\Delta t=\tau\left(  \frac{\chi_{s}+\bar{\chi}_{s}}{2}\right)  ,
\label{Del_t_tau}%
\end{equation}
where $\chi_{s}$ and $\bar{\chi}_{s}$ are launch angles of rays under
consideration and
\begin{equation}
\tau(\chi_{s})=2\pi I(\chi_{s})/c_{r}\text{.} \label{tau}%
\end{equation}
A solid line in Fig. 11 graphs $\tau(\chi_{s})$ for our model of
range-independent waveguide. Looking at this curve we can predict that, for
example, a difference in travel times of two eigenrays with launch angles
close to $7^{\circ}$ and numbers of cycle which differ by 1, will be close to
$\tau=0.19$ s at \textbf{any} range and at \textbf{any} depth, provided such
eigenrays arrive at the observation point.

\begin{figure}[ptb]
\begin{center}
\includegraphics[width=8cm,keepaspectratio=true]{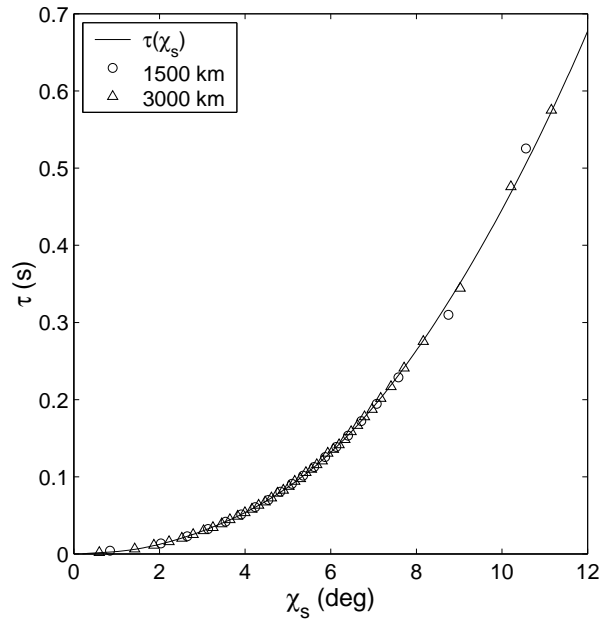}
\end{center}
\caption{The solid curve is $\tau=2\pi I(\chi)/c_{r}$ from Eq. (\ref{tau}).
The circles and triangles are time delays between segments with the
identifiers $+J$ and $+(J-2)$ at a reference depth of $0.78$ km computed for
even $J$ at 1500 km and 3000 km ranges, respectively.}%
\label{tauVSchi}%
\end{figure}

Let us select some reference depth, $z_{r}$, and define the temporal shift
between segments with identifiers $\pm J$ and $\pm(J-2)$ -- we denote this
shift by $T_{\pm J,\pm(J-2)}$ -- as a difference in travel times of two
eigenrays with these identifiers arriving at the depth $z_{r}$. Arrivals of
these rays in the upper panels of Fig. 2 and 3 can be found as intersections
of the corresponding segments with a horizontal line $z=z_{r}$. According to
Eq. (\ref{tau}) the value of $T_{\pm J,\pm(J-2)}$ does not depend on a
particular value of $z_{r}$ (the only requirement is that both segments must
intersect the line $z=z_{r}$). This result agrees with the fact that
neighboring segments in Figs. 2 and 3 with inclinations of the same sign are
almost parallel.

Since $T_{\pm J,\pm(J-2)}$ represent a difference in travel times of two
eigenrays, it can be estimated as $\tau(\chi_{\pm J,\pm(J-2)})$ with
$\chi_{\pm J,\pm(J-2)}$ being a half-sum of launch angles of the corresponding
eigenrays. It means that points depicting values of $T_{\pm J,\pm(J-2)}$
against $\chi_{\pm J,\pm(J-2)}$ should lie on the curve $\tau(\chi)$ at any
range and for any reference depth, $z_{r}$. In Fig. 11 this prediction is
verified for the timefront shown in the upper panels of Figs. 2 and 3 and for
a similar timefront at the range of 1500 km. Circles and triangles depict
$T_{+J,+(J-2)}$ as functions of $\chi_{+J,+(J-2)}$ at $1500$ km and $3000$ km
ranges, respectively, for $z_{r}=0.78$ km. Travel times shifts for only even
$J$ (from $60$ to $96$ at $1500$ km, and from $118 $ to $196$ at $3000$ km)
are shown. It is clearly seen that all the circles and triangles are, indeed,
located close to the solid curve representing $\tau(\chi)$.

\subsection{Timefront in the presence of perturbation}

\subsubsection{Widening and bias of timefront segments\label{widening}}

\bigskip In the presence of weak range-dependent inhomogeneities the structure
of timefront becomes more complicated: instead of infinitely thin segments of
smooth curves, we have some areas filled with randomly scattered points.
Although we observe the scattered points only because our fan is far too
sparse to resolve what should be unbroken curves, the appearance of such
regions indicates the presence of chaotic rays.

As it has been pointed out in Sec. \ref{timefront} the early portion of the
timefront formed by steep rays still ``remembers'' its structure in the
unperturbed waveguide. The points depicting arrivals of rays with the given
identifier are scattered in the vicinity of the corresponding unperturbed
segment. A group of arrivals formed by rays with the same identifier produces
a fuzzy versions of an unperturbed segment. We shall call such groups of
points in the time-depth plane, the fuzzy segments. Thus, every fuzzy segment,
like every segment of the unperturbed timefront, is associated with some
identifier. Two examples of fuzzy segments are shown in Figs. 2 and 3. We mean
two groups of thick points indicating arrivals of rays with the identifiers
$+124$ and $+160$. Closer views of these groups of points are shown in Fig.
12. The left (right) panel presents arrivals with the identifier $+124$
($+160$). Thick solid lines in both panels show the unperturbed segments with
the identifiers $+124$ (left panel) and $+160$ (right panel).

\begin{figure}[ptb]
\begin{center}
\includegraphics[width=7.5cm,height=7.5cm]{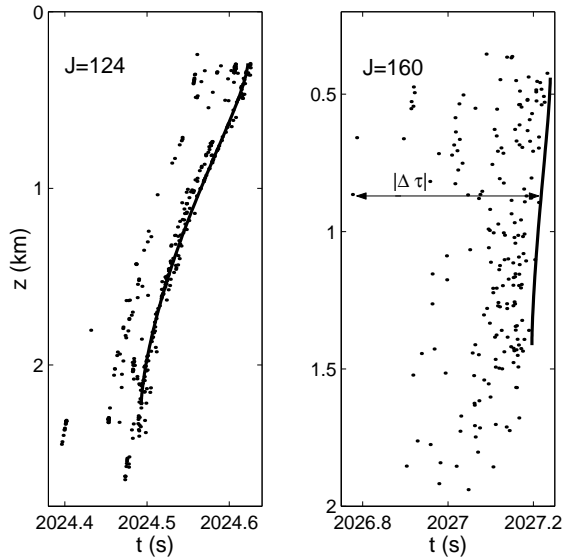}
\end{center}
\caption{Arrivals with identifiers $+124$ (left panel) and $+160$ (right
panel) at the range 3000 km presented in the time-depth plane. The points and
solid lines depict arrivals with and without internal waves present,
respectively. The magnitude of time delay, $\left|  \Delta\tau\right|  $,
between arrivals of the perturbed ray and an unperturbed ray with the same
identifier is shown for the earliest arrival with identifier $+160$. }%
\label{segments}%
\end{figure}

In order to derive quantitative characteristics of the fuzzy segment
describing its spread and bias, we introduce the quantity $\Delta\tau$ defined
as follows. Consider a particular ray (perturbed) contributing to the fuzzy
segment and denote its travel time by $t_{p}$. A travel time of an unperturbed
ray with the same identifier and the same arrival depth we denote by $t_{u}$.
Then
\begin{equation}
\Delta\tau=t_{p}-t_{u}. \label{Delta_tau}%
\end{equation}
In words, $\Delta\tau$ represents the distance along the $t$-axis between the
given point of the fuzzy segment and the unperturbed segment with the same
identifier. In the right panel of Fig. 12 the magnitude of $\Delta\tau$ is
shown for the earliest arrival. Note, that $\Delta\tau$ is defined only for
those rays forming the fuzzy segment whose arrival depths lie within a depth
interval covered by the unperturbed segment. \begin{figure}[ptb]
\begin{center}
\includegraphics[width=7.5cm,height=6.4cm]{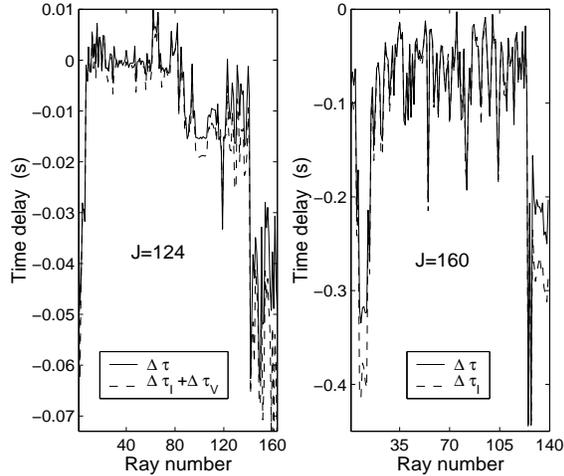}
\end{center}
\caption{Time delays, $\Delta\tau$, defined by Eq. (\ref{Delta_tau}), for
arrivals of rays with identifiers $+124$ (left panel) and $+160$ (right panel)
at the range 3000 km. The solid lines represent results of the direct ray
tracing, while the dashed lines are predictions made using Eq. (\ref{main}).
For rays with identifier $+160$ only the term $\Delta\tau_{I}$ in Eq.
(\ref{main}) has been taken into account.}%
\label{verify}%
\end{figure}An approximate analytical expression for $\Delta\tau$ is readily
obtained from Eq. (\ref{Del_t_short}). Since we consider eigenrays with
identical identifiers, the first two terms on the right of Eq.
(\ref{Del_t_short}) vanish and we arrive at
\begin{equation}
\Delta\tau=\Delta\tau_{I}+\Delta\tau_{V}, \label{main}%
\end{equation}
where
\begin{equation}
\Delta\tau_{I}=\frac{\bar{\omega}^{\prime}}{2c_{r}}\int\Delta I^{2}%
\,\,dr,\quad\label{Del_tau_I}%
\end{equation}
and
\begin{equation}
\Delta\tau_{V}=\frac{1}{c_{r}}\int\delta n(r,z(r))\,dr. \label{Del_tau_V}%
\end{equation}

First of all, check an accuracy of Eq. (\ref{main}) at $3000$ km range. This
is done in Fig. 13 where we present travel time shifts $\Delta\tau$ for
arrivals shown in Fig. 12. In the perturbed waveguide there are $160$ fan rays
with the identifier $+124$ whose depths at $3000$ km belong to the depth
interval covered by the unperturbed fan rays, i.e. between upper and lower
points of a solid curve in the left panel of Fig. 12. A solid line in the left
panel of Fig. 13 connects exact values of $\Delta\tau$ obtained by ray
tracing, while a dashed line represents predictions provided by Eq.
(\ref{main}). Launch angles of perturbed rays with the identifier $+124$
belong to the interval $(7^{\circ},9.5^{\circ})$. In Fig. 10 we see that the
parameter $\mu_{2}$ for such launch angles cannot be considered as small
compared to unity, which means that both terms in Eq. (\ref{main}) should be
retained. In the right panel of Fig. 13 a similar plot is shown for $\ 139$
rays with the identifier $+160$. The prediction depicted by a dashed curve has
been made by retaining only the term $\Delta\tau_{I}$, because launch angles
of these rays are less than $7^{\circ}$. The parameter $\mu_{2}$ for such
launch angles is small (see Fig. 10) and the term $\Delta\tau_{V}$ can be
neglected. Figure 13 demonstrates that Eq. (\ref{main}) provides a reasonable
estimation for $\Delta\tau$.

Figure 12 exhibits a new phenomenon. The perturbation causes not only
diffusion of the timefront segment but it also leads to some regular bias:
rays with the given identifier in the perturbed waveguide arrive, on average,
earlier than unperturbed rays with the same identifier. A similar bias is
observed for every fuzzy segment. A qualitative explanation to this effect
follows immediately from the fact that the term $\Delta\tau_{I}$ in Eq.
(\ref{main}) usually dominates. This is true even of steep rays although in
this case the term $\Delta\tau_{V}$ should be retained for obtaining an
accurate prediction of $\Delta\tau$. The sign of $\Delta\tau_{I}$ is
determined by the sign of the derivative $\bar{\omega}^{\prime}$. The latter
is negative for rays propagating without reflection off the surface and
bottom, because in typical deep ocean waveguides the cycle length of the
refracted ray grows with the launch angle. The spatial frequency $\omega$ and
its derivative with respect to $I$ are shown in Fig. 14 for the model of
unperturbed waveguide on which we rely in this paper.

\begin{figure}[ptb]
\begin{center}
\includegraphics[width=7cm,keepaspectratio=true]{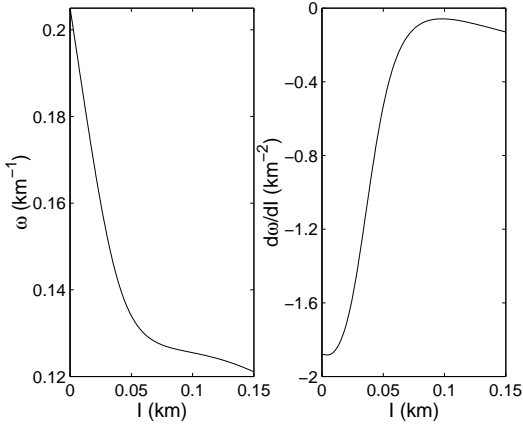}
\end{center}
\caption{Left panel: Spatial frequency of ray trajectory oscillation, $\omega
$, in the unperturbed waveguide as a function of the action variable, $I$.
Right panel: The derivative of the spatial frequency $\omega$ with respect to
$I$.}%
\label{OmdOm}%
\end{figure}

Approximating $\Delta\tau$ by $\Delta\tau_{I}$ (i.e., practically, by $\Delta
t_{I}^{(2)}$), one can estimate a mean bias of the fuzzy segment, $<\Delta
\tau>$, by making use of Eq. (\ref{Del_tI_aver}). In Fig. 15 we compare this
prediction (solid line) with \ ``real'' mean values of $\Delta\tau$ (circles)
evaluated for fuzzy segments formed by our fan rays with identifiers $+J$ at
ranges of 1500 km and 3000 km.

The rms time spread of the fuzzy segment can be estimated as
\[
\sigma_{\Delta\tau}=\sqrt{<(\Delta\tau-<\Delta\tau>)^{2}>}%
\]%
\[
=\sqrt{<\Delta\tau^{2}>-<\Delta\tau>^{2}}.
\]
Approximating again $\Delta\tau$ by $\Delta\tau_{I}$ and making use of Eq.
(\ref{M4}) we get
\[
<\Delta\tau^{2}>=\left(  \frac{\bar{\omega}^{\prime}}{2c_{r}}\right)  ^{2}%
\int_{0}^{r}\int_{0}^{r}d\rho_{1}d\rho_{2}%
\]%
\[
\times\left[  Q(\rho_{1},\rho_{1})Q(\rho_{2},\rho_{2})+2Q^{2}(\rho_{1}%
,\rho_{2})\right]  =\left(  \frac{\bar{\omega}^{\prime}}{2c_{r}}\right)
^{2}\frac{B^{2}r^{4}}{20}.
\]
This yields
\begin{equation}
\sigma_{\Delta\tau}=\frac{\left|  \bar{\omega}^{\prime}\right|  }{c_{r}}%
\frac{Br^{2}}{6\sqrt{5}}=0.9\,|<\Delta\tau>|, \label{sig_Del_tau}%
\end{equation}
i.e. the rms time spread is practically equal to its mean bias. In other
words, Eq. (\ref{Del_tI_aver}) estimates not only the bias of the fuzzy
segment but its time spread as well. In Fig. 16 we compare this prediction to
``real'' time spreads obtained in numerical simulation. Results presented in
Figs. 15 and 16 demonstrate that even though predictions made with Eq.
(\ref{Del_tI_aver}) have some systematic error for very flat rays
(corresponding to large $J$ in the lower panels of Figs. 15 and 16), these
predictions are in reasonable agreement with numerical simulation and give the
correct order of magnitude for both time spread and bias. In particular, we
see that at the 3000 km range, in accordance with Eq. (\ref{Del_tI_aver}), the
bias and time spread become 4 times larger compared to their values at 1500 km.

\begin{figure}[ptb]
\begin{center}
\includegraphics[width=7cm,keepaspectratio=true]{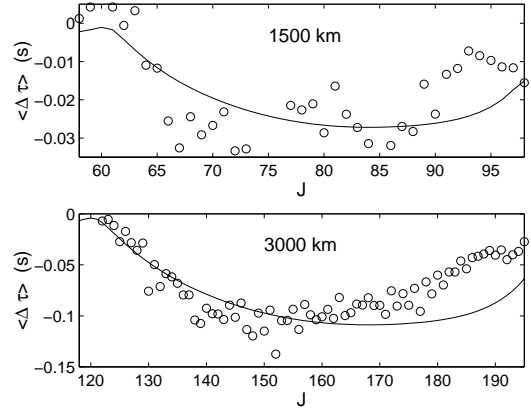}
\end{center}
\caption{Mean bias of fuzzy segments due to inhomogeneities as a function of
the ray identifier at ranges of 1500 km (upper panel) and 3000 km (lower
panel). Each circle shows the bias averaged over a fuzzy segment, i.e. over a
group of rays with the given identifier $+J$. The solid lines are predictions
obtained using Eq. (\ref{Del_tI_aver}).}%
\label{Bias}%
\end{figure}\begin{figure}[ptbptb]
\begin{center}
\includegraphics[width=7cm,keepaspectratio=true]{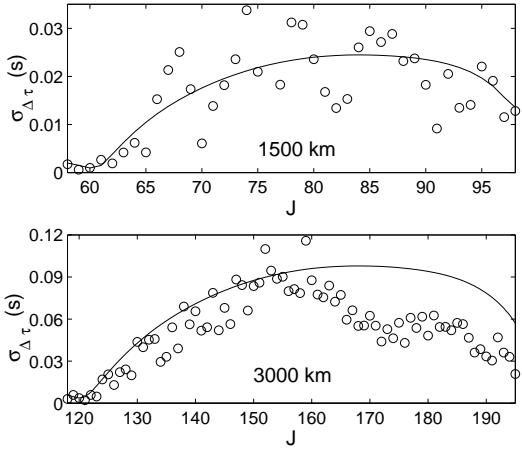}
\end{center}
\caption{Width of a fuzzy segment as a function of the ray identifier at
ranges of 1500 km (upper panel) and 3000 km (lower panel). Each circle shows
the standard deviation of the time delay $\tau$ obtained by averaging over the
fuzzy segment, i.e. over a group of rays with the given identifier $+J$. The
solid lines are predictions obtained using Eqs. (\ref{sig_Del_tau}) and
(\ref{Del_tI_aver}).}%
\label{Spread}%
\end{figure}

Looking at Figs. 15 and 16 we see that the bias and time spread are especially
small for segments formed by steep rays (small $J)$. On the right hand side of
Eq. (\ref{Del_tI_aver}) there is a factor $\bar{\omega}^{\prime}$, depending
on $\bar{I}$ and, hence, on the launch angle. Looking at Fig. 6 and at the
right panel in Fig. 14 we conclude that the dependencies of $<\Delta\tau>$ and
$\sigma_{\Delta\tau}$ on the launch angle are mainly determined by the factor
$\bar{\omega}^{\prime}$.

\subsubsection{Resolution of fuzzy segments in the perturbed timefront}

In the upper panels of Figs. 2 and 3 it is clearly seen that unperturbed
segments come in groups of four and each group consists of segments with the
identifiers $-(J-1)$, $\pm J$, and $+(J-1)$. The function $\tau(\chi)$ defined
by Eq. (\ref{tau}) predicts the time delay between two consecutive groups of
four formed by rays with launch angles close to $\chi$. The difference in
travel times of two neighboring segments can be roughly estimated as $\tau/4$.
Estimating the width of the fuzzy segment as $3\sigma_{\tau}$, we introduce
the parameter
\begin{equation}
R=\frac{12\sigma_{\tau}}{\tau}\simeq\frac{24\pi\bar{I}}{\left|  \bar{\omega
}^{\prime}\right|  Br^{2}} \label{Resolution}%
\end{equation}
representing the ratio of the segment width to the time delay between
neighboring segments. If $R>1$, the fuzzy segment is resolved in the perturbed
timefront while the segment with $R<1$ overlaps with neighboring segments.
From the viewpoint of eigenrays, the condition $R>1$ ($R<1$) means that
neighboring unperturbed eigenrays in the presence of perturbation split into
nonoverlapping (overlapping) clusters.

In order to verify this prediction, consider the ray travel time, $t$, in the
unperturbed waveguide as a function of the launch angle, $\chi_{s}$. For
launch angles of the same sign the function $t(\chi_{s})$ is monotonic and the
inverse function $\chi_{s}(t)$ is unambiguous. Equation (\ref{Resolution})
defines the parameter $R$ as a function of the launch angle $\chi_{s}$.
Replacing $\chi_{s}$ with $\chi_{s}(t)$ we obtain the function $R(t)$ that
associates a value of parameter $R$ with every segment of the unperturbed
waveguide. If this value is greater than unity we expect that the
corresponding fuzzy segment will be resolved. This statement is illustrated in
Figs. 17 and 18. The functions $R(t)$ at $r=1500$ km and $r=3000$ km evaluated
for rays starting upward (for rays starting downward \ the results are
practically the same) are plotted in the lower panels of Figs. 17 and 18.
These functions are shown at time intervals where $R(t)$ passes through $1$.
The corresponding portion of the perturbed timefronts from the upper panels of
Figs. 2 and 3 are shown in the upper panels of Figs. 17 and 18. It is clearly
seen that, indeed, the condition $R=1$ allows one to estimate the critical
travel time which divides the perturbed timefront into parts consisting of
resolved and unresolved fuzzy segments.

\begin{figure}[ptb]
\begin{center}
\includegraphics[width=7cm,keepaspectratio=true]{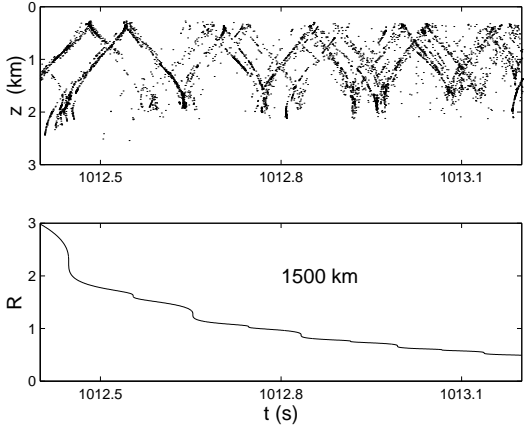}
\end{center}
\caption{Upper panel: Fragment of the timefront at 1500 km range where
transition from resolved to unresolved (overlapping) fuzzy segments is
observed. Lower panel: Parameter $R$ defined by Eq. (\ref{Resolution}) as a
function of travel time. Theory predicts that resolved (unresolved) fuzzy
segments are located to the left (right) of the travel time at which $R$
passes through $1$.}%
\label{R1500}%
\end{figure}\begin{figure}[ptbptb]
\begin{center}
\includegraphics[width=7cm,keepaspectratio=true]{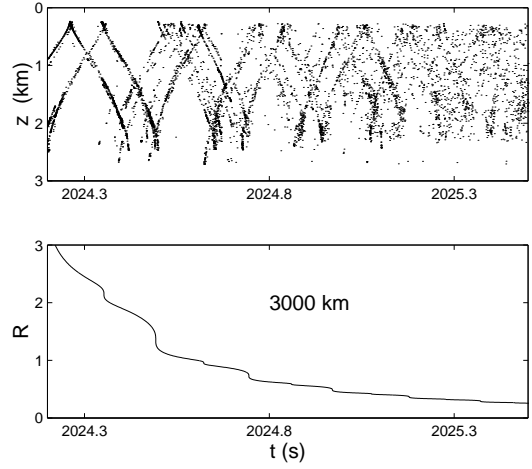}
\end{center}
\caption{The same as in Fig. 17 for 3000 km range.}%
\label{R3000}%
\end{figure}

The fact that early arriving fuzzy segments formed by steep rays are well
resolved is linked to the following two factors. First, the function
$\tau(\chi)$ grows with $|\chi|$. It means that the delay between two
consecutive segments of the timefront formed by steep rays is larger than that
for consecutive segments formed by flat rays. Second, the rms width of the
fuzzy segment, $\sigma_{\Delta\tau}$, is especially small for steep rays
(small $J$ 's), as it is seen in Fig. 16.

\subsubsection{Stability of fuzzy segments and Fermat's principle}

Probably, the most surprising feature of the perturbed timefront is
unexpectedly small widenings of timefront segments. In particular, in the
upper panel of Fig. 16 we see that at 3000 km range the maximum rms widening
of the fuzzy segment does not exceed 0.12 s. On the other hand, Fig. 4
demonstrates that a typical time spread for a cluster of rays with close
launch angles is about 2 s, i.e. much larger. Loosely, we can state that the
ray travel time dependence on the ray identifier is less chaotic and much more
predictable than its dependence on the launch angle.

In order to interpret this phenomenon, we come back to Eq. (\ref{main}) and
recall that the first term on the right side dominates. It means that the
travel time shift between perturbed ($P$) and unperturbed ($U$) rays
connecting the same points and having the \ same identifier can be
approximately written as
\begin{equation}
\Delta\tau=\frac{1}{c_{r}}[S_{0}(P)-S_{0}(U)] \label{dtI1}%
\end{equation}
where $S_{0}(P)$ and $S_{0}(U)$ are the values of the functional $S_{0}%
=\int(pdz-H_{0}dr)$ evaluated over the trajectories of perturbed and
unperturbed rays, respectively. According to Fermat's (Hamilton's) principle
\cite{LLmech,BornWolf,JKPS94}, the unperturbed ray provides a stationary path
of the functional $S_{0}$. This fact explains the absence of the linear in
$\Delta I$ term in Eqs. (\ref{Del_t_I}), (\ref{Del_t_short}), and Eq.
(\ref{main}). \ Since $\Delta I$ is our small parameter, the absence of terms
$O(\Delta I)$ gives some qualitative interpretation of smallness of
$\Delta\tau$. In this sense, the small time spread of clusters of rays with
the same identifier can be interpreted as a consequence of Fermat's principle.

Note that the difference in travel times of rays with different identifiers
($N\neq0$) is defined mainly by the term $\Delta t_{N}$ given by Eq.
(\ref{Del_t_N}) which usually is significantly larger than $\Delta t_{I}$.

\section{Action-angle variables in a waveguide with a range-dependent
background sound-speed profile\label{adiabatic}}

An ocean-acoustic propagation model with the sound speed field being a
superposition of a range-independent background and a weak range-dependent
perturbation responsible for emergence of ray chaos may be too idealized. In
this section we shortly outline how the results obtained in the preceding
sections can be generalized to a more realistic model.

First, let us shortly discuss a method of introducing of the action-angle
variables in the range-dependent waveguide \cite{LLmech} without dividing the
Hamiltonian into a sum of an unperturbed term and a perturbation. Define
canonical transformations (\ref{canonH}) and (\ref{canoninvH}) at a current
range $r$ using Eqs. (\ref{canon1H}) and (\ref{canon2H}) evaluated for an
auxiliary range-independent waveguide with the same cross-section that the
real waveguide has at the range $r$. In this case the canonical transformation
will be different at different ranges and Eqs. (\ref{canonH}) and
(\ref{canoninvH}) translate to
\begin{equation}
I=I(p,z,r),\,\,\,\,\,\theta=\theta(p,z,r) \label{canonH1}%
\end{equation}
and
\begin{equation}
z=z(I,\theta,r),\quad p=p(I,\theta,r). \label{canoninvH1}%
\end{equation}
The generating function $G$ now becomes a function of not only $I$ and $z$,
but of $r$, as well. However, $H=-\sqrt{n^{2}-p^{2}}$ in the new variables is
a function of $I$ and $r$, but not $\theta$ \cite{LLmech}.

The Hamilton equations in the new variables preserve their canonical form
\begin{equation}
\frac{dI}{dr}=-\frac{\partial H_{s}}{\partial\theta},\quad\frac{d\theta}%
{dr}=\frac{\partial H_{s}}{\partial I} \label{dIdr0H1}%
\end{equation}
with the new Hamiltonian \cite{LLmech}%

\begin{equation}
H_{s}(I,\theta,r)=H(I,r)+\Lambda(I,\theta,r),\label{H1H}%
\end{equation}
where
\begin{equation}
\Lambda(I,\theta,r)=\left.  \frac{\partial G(I,z,r)}{\partial r}\right|
_{z=z(I,\theta,r)}.\label{LambdaH}%
\end{equation}

The term $\Lambda$ is small and can be neglected if range variations in the
environment are adiabatic, i.e. if variations in the environment are small at
the cycle of the ray trajectory. Then, $dI/dr=0$ and $I$ remains constant
along the ray trajectory, i.e. the action variable defined in this way does
have a property of adiabatic invariance.

However, we suppose that this common approach is not convenient for
description of the chaotic ray motion induced by random internal waves. The
point is that if $\Lambda$ is not negligible, then the connection between $H
$, $\Lambda$ and $\delta c$ becomes non-trivial, and it is difficult to divide
Hamiltonian (\ref{H1H}) into a sum of a smooth unperturbed term and a small
perturbation. But such a decomposition of the Hamiltonian is necessary for
application of our perturbation theory.

A more appropriate approach can be developed if the sound speed field is a sum
of a \textbf{smooth} range-dependent sound speed, $c_{0}(r,z)$, and a weak
perturbation, $\delta c(r,z)$, i.e.
\begin{equation}
c(r,z)=c_{0}(r,z)+\delta c(r,z). \label{crz1}%
\end{equation}
Instead of the range-independent unperturbed waveguide considered in the
preceding sections, now we have an adiabatic one. Then, it is convenient to
introduce the action-angle variables at every range $r$ using an auxiliary
range-independent waveguide with the cross-section coinciding with that of the
unperturbed waveguide. This yields the new Hamiltonian in the form%

\begin{equation}
H_{s}=H_{0}(I,r)+V(I,\theta,r), \label{HsH1}%
\end{equation}
where $V(I,\theta,r)$ is the perturbation defined in Eq. (\ref{H0V}) with the
unperturbed refractive index $n_{0}$ now depending not only on $z$ but on $r$,
as well. Then the Hamilton equations have the same form as Eqs. (\ref{dIdrH})
and (\ref{dtedrH}) in Sec. \ref{range-dep}, although the angular frequency
$\omega$ now depends on $r$, $\omega(I,r)=\partial H_{0}(I,r)/\partial I$. All
expressions for differences in ray travel times of perturbed and unperturbed
rays derived in Sec. \ref{travel} remain valid for this more realistic model.

\section{Summary and conclusion\label{conclusion}}

\bigskip In this paper we have derived the approximate analytical approach for
description of ray travel times and other parameters of the ray structure in
deep ocean environment. Our results remain valid at ranges up to, at least, a
few thousand km. The approach is based on the assumptions that (i) the
perturbation giving rise to the chaotic ray motion is small and (ii) even at
long ranges rms variations of the action variable are small compared to the
characteristic scale of function $\omega(I)$. The dimensionless small
parameters in the problem are given by Eqs. (\ref{mu1}) and (\ref{mu2}).

\bigskip The exact expression for the difference in travel times of perturbed
and unperturbed rays, $\Delta t$, determined by Eqs. (\ref{constituents}) --
(\ref{Del_t_V}) has been significantly simplified by expanding it in a power
series in $\Delta I$ and neglecting small terms. The smallness of fluctuations
of the action variable has also been used to simplify the stochastic ray
(Hamilton) equations. It has been shown that the fluctuating components of the
action and angle variables can be idealized as a Wiener process, and an
integral of the Wiener process, respectively. This result, which allows one to
evaluate (approximately) practically any statistical characteristic of the ray
trajectory, in the present paper has been combined with an approximate formula
for $\Delta t$ and applied for investigation of range variations of ray travel times.

Our primary concern has been with the range variations of the timefront
representing ray arrivals in the time-depth plane. The unperturbed timefront
consists of segments of smooth curves. Each segments is formed by rays with
the same identifier. In the presence of perturbation segments become fuzzy:
arrivals with the given identifier form a set of points randomly scattered
around the unperturbed segment. The time spread of these points turns out to
be unexpectedly small. It is much less than a time spread of arrivals with
launch angles within a narrow angular interval corresponding to launch angles
of rays forming an unperturbed segment. The most apparent manifestation of
this phenomenon is a surprising stability of early portions of the timefront
formed by steep rays \cite{BV98,SFW97,AET1,AET2}.

Our approach provides a quantitative description of fuzzy segments. It gives
estimations of their widths and biases. Using these estimations, it follows
that the sign of the bias is determined by the sign of $d\omega/dI$, the
derivative of the spatial frequency of ray oscillation in the unperturbed
waveguide with respect to the action variable. In typical deep ocean
waveguides $d\omega/dI<0$ for all refracted rays, which means that fuzzy
segments have negative bias, i.e. perturbed rays, on average, arrive earlier
compared to unperturbed rays with the same identifier. It has been shown that
the surprising stability of fuzzy segments with respect to the perturbation is
related to the Fermat's (Hamilton's) principle.

The estimations derived for the timefront segments can be applied to study
characteristics of chaotic eigenrays. In Ref. \cite{TT96} (see also Ref.
\cite{ST96}) it has been discovered numerically, that in the presence of a
weak perturbation the unperturbed eigenray splits into a cluster of new
eigenrays with close arrival times. The time spread of such a cluster can be
estimated as a time spread of the corresponding fuzzy segment at the
corresponding depth. So, our criterion of nonoverlapping of neighboring
segments (\ref{Resolution}) provides the criterion of resolution for clusters
of eigenrays.

It should be emphasized that results of the present work have been obtained
for the perturbation induced by internal waves. Their generalization to the
case of other inhomogeneities, e.g. the mesoscale inhomogeneities, requires a
further investigation.

\section*{Acknowledgment}

I thank Prof. G. Zaslavsky for benefit of our discussions on problems of
quantum and wave chaos. I am grateful to Dr. I.P. Smirnov whose ray code has
been used for numerical simulations. The work was supported by the U.S. Navy
Grant N00014-97-1-0426 and by the Grants No. 00-02-17409 and 01-05-64394 from
the Russian Foundation for Basic Research.

\appendix

\section{Appendix. Evaluation of action and angle variables using a ray
code\label{evaluation}}

Explicit analytical relations connecting the position-momentum and
action-angle variables can be derived only for a few simple refraction index
profiles $n_{0}(z)$ \cite{AZ91,Abdullaev}. In underwater acoustics the
researcher is usually forced to deal with refraction index profiles defined by
spline approximations of data measured at discrete depths. In this case ray
trajectories are computed using standard ray codes which allow one evaluate
such parameters of the ray as the coordinate, momentum, and travel time
\cite{JKPS94}. Fortunately, the same codes can be applied for evaluation of
the action and angle variables. It will be a simpler and more convenient
procedure compared to the direct application of Eqs. (\ref{canon1H}) and
(\ref{canon2H}).

As it has been pointed out in Secs. \ref{range-dep} and \ref{adiabatic} the
transformation from the position-momentum to the action-angle variables is
\textbf{always }defined using an auxiliary range-independent waveguide. So
what we need for establishing relations between the two types of variables in
any waveguide, is a code for computing the ray path in the range-independent waveguide.

The action variable for the given trajectory can be easily found by evaluating
the integral
\begin{equation}
I=\frac{1}{2\pi}\int_{0}^{D}\frac{p^{2}(r)}{\sqrt{n_{0}^{2}(z(r))-p^{2}(r)}}dr
\label{I_numeric}%
\end{equation}
where the functions $z(r)$ and $p(r)$ are the coordinate and the momentum,
respectively, as functions of range, and $D$ is the cycle length. Here we have
applied Eqs. (\ref{IH}) and (\ref{dzdrH}). Computing ray paths corresponding
to different values of the Hamiltonian (\ref{SnellH}) and using a spline
approximation we find functions $H_{0}(I)$, $I(H_{0})$, and $\omega
(I)=2\pi/D(I)$.

Then the transformation from $(p,z)$ to $(I,\theta)$ is performed as follows.
Substituting the given $p$ and $z$ into Eq. (\ref{SnellH}) yields $H_{0}$.
Then, making use of the function $I(H_{0})$ we get the action variable, $I$.
In order to find the angle variable, $\theta$, consider a ray trajectory with
$H_{0}$ determined by the given $p$ and $z$, which begins at its upper turning
point, i.e. at $z(0)=z_{\min}$ (recall that we are using $z$-axis directed
downward). According to Eq. (\ref{I0H})
\begin{equation}
r=\theta/\omega(I). \label{r_theta}%
\end{equation}
It means that a minimum range, $r$, at which the trajectory has the coordinate
and momentum equal to the given $p$ and $z$, respectively, defines the desired
angle variable, $\theta$.

In order to use Hamilton equations (\ref{dIdrH}) and (\ref{dtedrH}) in
practice we need a numerical procedure which allows one to express an
arbitrary function $F(p,z)$ as a function of $I$ and $\theta$, by replacing
the arguments $p$ and $z$ with $p(I,\theta)$ and $z(I,\theta)$. In other
words, we need a function $\Phi(I,\theta)\equiv F(p(I,\theta),z(I,\theta))$.
In principle, it can be found by a direct substitution. However, there exists
a following more convenient procedure.

Since both $p(I,\theta)$ and $z(I,\theta)$ are periodic in $\theta$ with the
period $2\pi$, the function $\Phi(I,\theta)$ is also periodic and it can be
represented as a Fourier series
\begin{equation}
\Phi(I,\theta)=\sum_{n=0}^{\infty}A_{n}(I)\cos n\theta+\sum_{n=1}^{\infty
}B_{n}(I)\sin n\theta, \label{Phi_I_T}%
\end{equation}
where
\[
\left(
\begin{array}
[c]{l}%
A_{n}(I)\\
B_{n}(I)
\end{array}
\right)  =M_{n}%
\]%
\begin{equation}
\times\int_{0}^{2\pi}d\theta\,\Phi(I,\theta)\left(
\begin{array}
[c]{l}%
\cos n\theta\\
\sin n\theta
\end{array}
\right)  , \label{A_B_T}%
\end{equation}%
\[
\;M_{n}=\left\{
\begin{array}
[c]{l}%
1/2\pi,\;n=0\\
1/\pi,\;\;n>0
\end{array}
\right.  .
\]

Using relation (\ref{r_theta}) we can express the above integrals via the
integrals over the ray trajectory computed with the ray code. This yields
\[
\left(
\begin{array}
[c]{l}%
A_{n}(I)\\
B_{n}(I)
\end{array}
\right)  =M_{n}\omega(I)
\]%
\begin{equation}
\times\int_{0}^{2\pi}dr^{\prime}\,F(p(r^{\prime}),z(r^{\prime}))\left(
\begin{array}
[c]{l}%
\cos n\omega(I)r^{\prime}\\
\sin n\omega(I)r^{\prime}%
\end{array}
\right)  . \label{A_B_r}%
\end{equation}
If the function $F$ depends not only on $p$ and $z$, but on range $r$, as
well, then the function $\Phi$ and the coefficients $A_{n}$ and $B_{n}$
acquire an additional argument $r$. In Eq. (\ref{A_B_r}) this argument should
be considered as a constant, i.e. there should be no integration over this argument.

Note that functions $p(r)$ and $z(r)$ in Eq. (\ref{A_B_r}) which define a
trajectory in the auxiliary waveguide may be quite different from range
dependencies of coordinates and momenta for ``real'' ray trajectories
satisfying Hamilton equations (\ref{dzdrH}) and (\ref{dpdrH}).

%
%

\end{document}